\documentclass[journal]{IEEEtran}
\usepackage{graphicx}
\usepackage{amsmath}
\usepackage{amssymb}
\usepackage{subfigure}
\usepackage{tabu}
\usepackage{cite}
\usepackage{indentfirst}
\usepackage[utf8]{inputenc}
\usepackage[english]{babel}
\usepackage{bbm} 

\usepackage{epstopdf}
\usepackage{amsxtra}
\usepackage{amstext}
\usepackage{amsthm}
\usepackage{latexsym}
\usepackage{dsfont} 
\usepackage{xcolor}
\usepackage{units}
\usepackage{multirow}
\usepackage{lipsum}
\usepackage{bm}
\usepackage{multicol}
\usepackage{setspace}

\definecolor{darkgreen}{rgb}{0,0.75,0}
\usepackage[colorlinks=true]{hyperref}
\hypersetup{
	pdfauthor = {Asmaa W. Abdallah},
	pdfcreator = {Asmaa W. Abdallah},
	pdfnewwindow,
	pdffitwindow=true,
	colorlinks=true,
	linkcolor=blue,       
	citecolor=darkgreen,      
	filecolor=red,        
	urlcolor=cyan
}

\makeatletter
\let\NAT@parse\undefined
\makeatother

\theoremstyle{remark}

\theoremstyle{definition}

\usepackage{algorithm}
\usepackage[noend]{algpseudocode}
\algnewcommand{\LineComment}[1]{\State \(\triangleright\) #1}

\newcommand{\nth}[1]{{#1}{\text{th}}}
\newcommand{\mbf}[1]{\mathbf{#1}}

\newcommand{\abs}[1]{\left|{#1}\right|}
\newcommand{\norm}[1]{\left\|{#1}\right\|}

\newcommand{\expec}[1]{{\mathbb{E}\!\left[{#1}\right]}}

\begin{document}

	\title{Efficient Angle-Domain Processing for FDD-based Cell-free Massive MIMO Systems}
	
	\author
	{
		Asmaa~Abdallah,~\IEEEmembership{Student Member,~IEEE,}
		and Mohammad~M.~Mansour,~\IEEEmembership{Senior Member,~IEEE}
		
		\thanks{
			Part of this work has been presented at the 2019 IEEE International Workshop on Signal Processing Advances in Wireless Communications (SPAWC)~\cite{Asmaa2019cellfree}.
		}
		
		\thanks{
			This work is supported by the National Council for Scientific Research of the Lebanese Republic (CNRS-L) and the American University of Beirut (AUB) fellowship program, as well as by the University Research Board (URB) at AUB.
		}
		
		\thanks{
			A. Abdallah and M. M. Mansour are with the Department of Electrical and Computer Engineering, American University of Beirut, Beirut 1107 2020, Lebanon (e-mail: awa18@aub.edu.lb; mmansour@aub.edu.lb).
		}
		
	}
	
	\vspace{-0.2in}	
	
	\maketitle
	
	\begin{abstract}
		Cell-free massive MIMO communications is an emerging network technology for 5G wireless communications wherein distributed multi-antenna access points (APs) serve many users simultaneously. Most prior work on cell-free massive MIMO systems assume time-division duplexing mode, although frequency-division duplexing (FDD) systems dominate current wireless standards. The key challenges in FDD massive MIMO systems are channel-state information (CSI) acquisition and feedback overhead. To address these challenges, we exploit the so-called angle reciprocity of multipath components in the uplink and downlink, so that the required CSI acquisition overhead scales only with the number of served users, and not the number of AP antennas nor APs. We propose a low complexity multipath component estimation technique and present linear angle-of-arrival (AoA)-based beamforming/combining schemes for FDD-based cell-free massive MIMO systems. We analyze the performance of these schemes by deriving closed-form expressions for the mean-square-error of the estimated multipath components, as well as expressions for the uplink and downlink spectral efficiency. Using semi-definite programming, we solve a max-min power allocation problem that maximizes the minimum user rate under per-user power constraints.  {Furthermore, we present a user-centric (UC) AP selection scheme in which each user chooses a subset of APs to improve the overall energy efficiency of the system.} Simulation results demonstrate that the proposed multipath component estimation technique outperforms conventional subspace-based and gradient-descent based techniques. We also show that the proposed beamforming and combining techniques along with the proposed power control scheme substantially enhance the spectral and energy efficiencies with an adequate number of antennas at the APs.
	\end{abstract}
	

	\section{Introduction}
	With the growing demand on high data-rate wireless communications, fifth generation (5G) cellular mobile communications has emerged as the latest generation to offer 1000-fold capacity enhancement over current fourth generation (4G) Long-Term Evolution (LTE) systems with reduced latency. To achieve this aggressive goal, massive multiple-input multiple-output (MIMO) and network densification are promising 5G wireless technologies that improve the capacity of cellular systems by 1) scaling  up the number of antennas in a conventional MIMO system by orders of magnitude~\cite{Marzetta2010MIMO,Marzetta2014mimo}, and 2) reducing path-loss and reusing spectrum~\cite{Ji2017OverviewMIMOLTE} efficiently.
	
	Although massive MIMO and network densification bring forward several advantages, the performance of cellular networks is limited by inter-cell interference (ICI) and frequent handovers for fast moving users. In particular, users close to the cell edge suffer from strong interference.
	
	Cell-free (CF) massive MIMO has recently been considered as a practical and useful embodiment of network MIMO that can potentially reduce such inter-cell interference through coherent cooperation between base stations~\cite{Nayebi2015CFmMIMO,Ngo2017CellFreevsSmallCell,Ngo2018EEofCF,interdonato2018ubiquitous}. In cell-free massive MIMO, the serving antennas are distributed over a large area. Distributed systems can potentially provide higher coverage probability than co-located massive MIMO due to their ability to efficiently exploit diversity against shadow fading effects, at the cost of increased backhaul requirements~\cite{Zhou2003Dist}.
	
	According to~\cite{interdonato2018ubiquitous}, ``cell-free'' massive MIMO implies that, from a user perspective during data transmission, all access points (APs) cooperate to jointly serve the end-users; hence there are no cell boundaries and no inter-cell interference in the data transmission. The APs are connected to a central processing unit (CPU) via a backhaul link. This approach, with simple signal processing, can effectively control ICI, leading to significant improvements in spectral and energy efficiency over the cellular systems~\cite{Ngo2017CellFreevsSmallCell,Ngo2018EEofCF,interdonato2018ubiquitous,Zhou2003Dist,Nayebi2015CFmMIMO}.

	The main challenge in deploying cell-free networks lies mainly in acquiring sufficiently accurate channel state information (CSI) so that the APs can simultaneously transmit (receive) signals to (from) all user equipments (UEs) and cancel interference in the spatial domain. The conventional approach of sending downlink (DL) pilots and letting the UEs feed back channel estimates is unscalable since the feedback load is proportional to the number of APs. Therefore, to reduce the signaling overhead~\cite{Emil2010coop,marzetta2016fundamentals}, channel reciprocity can be exploited in time-division duplex (TDD) mode so that each AP only needs to estimate the uplink CSI.
	
	An attractive alternative to consider is frequency-division duplexing (FDD) based cell-free massive MIMO systems for the following reasons: 1) channel reciprocity in TDD mode might not be accurate due to calibration errors in radio frequency (RF) chains \cite{Vieira2017calibration}, 2) with the lack of downlink training symbols in TDD systems, users may not be able to acquire instantaneous CSI, and thus system performance will deteriorate in detecting and decoding the intended signals, 3) while TDD operation is preferable at sub-6 GHz massive MIMO, in millimeter wave (mmWave) bands FDD may be equally good since the angular parameters of the channel are reciprocal over a wide bandwidth~\cite{bjornson2018massive}, and 4) FDD systems dominate current wireless communications and have many benefits such as lower cost and greater coverage than TDD~\cite{qualcommFDDTDD}.
	
	On the other hand, FDD-based cell-free massive MIMO systems still suffer from CSI acquisition and feedback overhead since the amount of downlink CSI feedback scales linearly with the number of antennas~\cite{Lee2015FDDmMIMO} and the number of APs in cell-free massive MIMO system. However, we can still benefit from 1) angle reciprocity, which holds true for FDD systems as long as the uplink and downlink carrier frequencies are not too far from each other (less than several GHz~\cite{Gao2017UnifiedTDDFDD}), and 2) angle coherence time which is much longer than the conventional channel coherence time~\cite{Heath2017Beamwidth} where the channel angle information can be regarded as unchanged. Hence, angle information is essential in FDD-based cell-free massive MIMO systems. Therefore, a low complexity estimation approach that can efficiently estimate the angle information is required.

	\subsection{Related Work}
	Much of the recent interest in cell-free massive MIMO systems has focused mainly on TDD-mode only~\cite{Ngo2017CellFreevsSmallCell,Ngo2018EEofCF,interdonato2018ubiquitous,Zhou2003Dist,Nayebi2015CFmMIMO,Alonzo2017CFandUCmmWave,Alonzo2019CF,maxminPCUL2019Bashar,Ngo2018CF}. In~\cite{Nayebi2015CFmMIMO}, a cell-free system is considered and algorithms for power optimization and linear precoding are analyzed. Compared with the conventional small-cell scheme, cell-free massive MIMO can yield more than ten-fold improvement in terms of outage rate. While in~\cite{Ngo2017CellFreevsSmallCell}, the APs perform multiplexing/de-multiplexing through conjugate beamforming in the downlink and matched filtering in the uplink.
	
	In~\cite{Ngo2018EEofCF}, a cell-free massive MIMO downlink is considered, wherein a large number of distributed multiple-antenna APs serve many single-antenna users. A distributed conjugate beamforming scheme is applied at each AP via the use of local CSI. Spectral efficiency and energy efficiency are studied while considering channel estimation error and power control.
	
	{In~\cite{Alonzo2017CFandUCmmWave,Alonzo2019CF}, cell-free and user-centric architectures at mmWave frequencies are considered. A multiuser clustered channel model is introduced, and an uplink multiuser channel estimation scheme is described along with hybrid analog/digital beamforming architectures. Moreover, in \cite{Alonzo2019CF}, the non-convex problem of power allocation for downlink global energy efficiency maximization is addressed. }	In~\cite{maxminPCUL2019Bashar}, an uplink TDD-based cell-free massive MIMO system is considered. Geometric programming {GP} is used to sub-optimally solve a quasi-linear max–min signal-to-interference-and-noise ratio (SINR) problem.

	Angle estimation has been studied in other wireless networks without considering cell-free massive MIMO networks (see e.g. ~\cite{Shmidt1986MUSIC1,Roy1989ESPIRIT,Krim1996ASP,Wang2015DOAmMIMO,Gao2014ESPIRIT,Cheng2015DoA,Shafin2016DoAmmwave,Gao2017DoAestMmwave,Gao2017UnifiedTDDFDD,GAO2018AODest }). For instance, subspace-based angle estimation algorithms, such as multiple signal classification (MUSIC), estimation of signal parameters via rotational invariance technique (ESPRIT) and their extensions have gained interest in the array processing community due to their high resolution angle estimation capability~\cite{Shmidt1986MUSIC1,Roy1989ESPIRIT,Krim1996ASP}. Their applications in massive MIMO systems and MIMO systems for angle estimation have been presented in~\cite{Wang2015DOAmMIMO,Gao2014ESPIRIT,Cheng2015DoA,Shafin2016DoAmmwave}. Unfortunately, the classical MUSIC and ESPRIT schemes are not suitable for mmWave communications due to the following main reasons: 1) They have high computational complexity mainly due to the singular value decomposition (SVD) operation on channels with massive number of antennas; 2) They are considered as blind estimation techniques originally targeted for radar applications, and do not make full use of training sequences in wireless communication systems.
	
	In~\cite{Gao2017DoAestMmwave,Gao2017UnifiedTDDFDD,GAO2018AODest }, an AoA estimation scheme for a conventional mmWave massive MIMO system with a uniform planar array at the base station is presented. The initial AoAs of each uplink path are estimated through the two-dimensional discrete Fourier transform (2D-DFT), and then the estimation accuracy is further enhanced via an angle rotation technique. In the present work, we extend the AoA estimation technique of~\cite{Gao2017DoAestMmwave,Gao2017UnifiedTDDFDD,GAO2018AODest}, adapt it to the context of FDD-based cell-free massive MIMO, and employ it to estimate another channel multipath component, namely large-scale fading. Using these estimated components, we leverage from the angle coherence time and angle-reciprocity to propose low-complexity angle-based beamforming/combining schemes and power control algorithms for downlink and uplink directions.

	In~\cite{Kim2018CFreeFDD}, a multipath component estimation technique and base station cooperation scheme based on the multipath components for the FDD-based cell-free massive MIMO systems are presented. However, no closed-form expression of the mean-square-error (MSE) of the considered multipath estimation is presented.

	\subsection{Contributions of the Paper}
	In this work, we consider a cell-free massive MIMO system with multiple antennas at each AP operating in FDD mode that do not require any feedback from the user. All APs cooperate via a backhaul network to jointly transmit signals to all users in the same time-frequency resources. By exploiting angle reciprocity, APs can acquire multipath component information from the uplink pilot signals using array signal processing techniques. The contributions of this paper are:
	\begin{enumerate}
		\item We propose a multipath component estimation for the AoA and large-scale fading coefficients based on the DFT operation and log-likelihood function with reduced overhead. In particular, we leverage from the observation that the angle-of-departures (AoDs) and the large scale fading components vary more slowly than path gains~\cite{Heath2017Beamwidth}, as well as from the property of angle-reciprocity. We further derive a closed-form expression for the MSE of the estimated channel multipath components. 
		Both theoretical and numerical results are provided to verify the effectiveness of the proposed methods. These schemes are shown to provide a substantial enhancement over the gradient-based~\cite{Kim2018CFreeFDD} and the classical subspace-based~\cite{Shmidt1986MUSIC1,Roy1989ESPIRIT} multipath component estimation in terms of MSE of the estimated AoA and large-scale fading coefficients since the MSE of the proposed DFT-based estimator coincides with that of the ML estimator.
		\item We propose linear angle-based beamforming/combining techniques for the downlink/uplink transmission that incorporate the estimated AoA and large-scale fading components. Interestingly, the proposed schemes scale only with the number of served users rather than the total number of serving antennas, and need to be updated every angle coherence time. Therefore, the impact of signaling overhead is substantially reduced with the proposed schemes.
		\item We derive closed-form expressions for the spectral efficiencies for the FDD-based cell-free massive MIMO downlink and uplink with finite numbers of APs and users. Our analysis takes into account the proposed beamforming/combining techniques and the effect of multipath estimation errors.
		\item {We propose a solution to the max-min power control problem by formulating it as a standard semi-definite programming (SDP) approach. The proposed max-min power control maximizes the smallest rate of all users within the angle-coherence time-scale. In addition, we present a user-centric AP selection scheme to further enhance the energy efficiency of the system.}
	\end{enumerate}
	
	The rest of the paper is organized as follows. The system model for the FDD-based cell-free massive MIMO network is described in Section~\ref{s:System_Model}. In Section~\ref{s:angle_estimation}, the proposed multipath components estimation is introduced. In Section~\ref{s:Beam_Com}, the proposed beamforming and combining techniques are presented. Moreover, spectral efficiency analysis is introduced in Section~\ref{s:Spectral_efficiency_Analysis}. Case studies with numerical results are simulated and analyzed based on the proposed schemes in Section~\ref{sec:simulation}. Section~\ref{sec:conclusion} concludes the paper.
	
	\emph{Notation}: Bold upper case, bold lower case, and lower case letters correspond to matrices, vectors, and scalars, respectively. Scalar norms, vector $\text{L}_2$ norms, and Frobenius norms, are denoted by $\abs{\cdot}$, $\norm{\cdot}$, and $\norm{\cdot}_{\text{F}}$, respectively. $\expec{\cdot}$, $(\cdot)^{\mathsf{T}}$, $(\cdot)^*$, $(\cdot)^{\mathsf{H}}$,  $\mbf{P}^{\perp}$, and  $\text{tr} (\cdot)$  stand for expected value, transpose, complex conjugate, Hermitian,  orthogonal projection matrix, and the trace of a matrix. {$\bf{X}^\dagger $ stands for the pseudo-inverse $(\bf{X}^\mathsf{H}\bf{X})^{-1}\bf{X}^\mathsf{H}$.} In addition, $\bf X \succeq 0$ is used to indicate that $\mbf{X}$ is a positive semi-definite matrix. $[\mbf{x}]_i$ represents $\nth{i}$ element of a vector $\mbf{x}$. $\mathcal{CN}(0,\sigma_{n}^{2})$ refers to a circularly-symmetric complex Gaussian distribution with zero mean and variance $\sigma_{n}^{2}$.
	\begin{figure}[t]
		\centering		\includegraphics[scale=0.33]{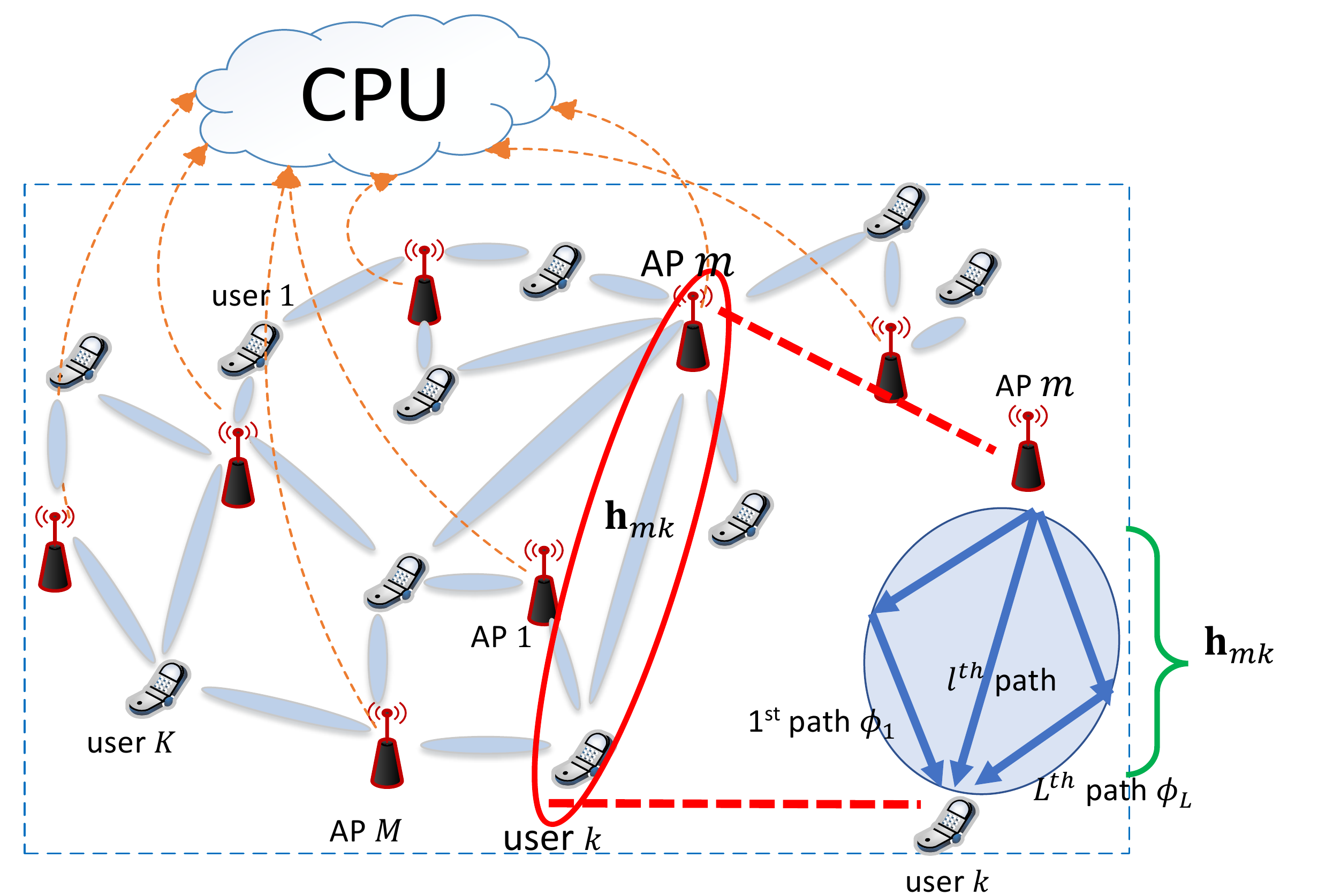}\vspace{-0.15in}
		\caption{Cell-free massive MIMO system model}\label{fig:model}
	\end{figure}
	
	\section{System Model}\label{s:System_Model}
	{As shown in Fig.~\ref{fig:model}, we consider an FDD-based cell-free massive MIMO system having $M$ APs, each equipped with a uniform linear array (ULA) of $N$ antennas, serving $K$ users with single antennas. We assume a geometric channel model with $L$ propagation paths \cite{Gao2017UnifiedTDDFDD,Kim2018CFreeFDD}. Moreover, AoAs (or AoDs), large-scale fading and small-scale fading coefficients are called the multipath components of the channel. Due to angle reciprocity in FDD systems ~\cite{Gao2017UnifiedTDDFDD}, and frequency in-dependency, we assume that 1) the uplink AoA and downlink AoD are similar, and 2)  the uplink and downlink  large-scale fading coefficients (slow fading and distant-dependent path loss components) are similar \cite{Marzetta2013ICILargescale,Larsson2018FDDperformance}. However, uplink and downlink small-scale fading coefficients in FDD systems are distinct since they are frequency dependent \cite{Marzetta2013ICILargescale,Larsson2018FDDperformance}.   Therefore, the $N\times 1$ channel vectors can be expressed as \cite{Gao2017UnifiedTDDFDD,Kim2018CFreeFDD}
		\begin{align}\label{channel_model}
		\bf{h}_{}&=\sqrt { \frac{1}{ L} } \: \sum_{l=1}^{L} \sqrt{\beta_{l}}\alpha_{l}\textbf{a}(\phi_{l}),
		\end{align}
		where $\alpha_{l} \sim \mathcal{CN}(0,1)$ is the complex gain of the $\nth{l}$ path that represents the small-scale {Rayleigh} fading, and $\beta_{l}$ is the large-scale fading coefficient that accounts for path-loss and shadowing effects. The variable $\phi_{l} \in [0, 2\pi]$ is the angle of arrival of the $\nth{l}$ path. }The array steering vector ${\bf a}\left (\phi_{l} \right)$ is defined as $	{\bf a}\left (\phi _{l} \right)\!= \!{\tfrac{1}{\sqrt {N}}}\!\! \left[1, e^{j {\eta \sin \left (\phi _{l} \right) } }\!,\!  \ldots,\! e^{j\left (N -1\right){\eta \sin \left (\phi _{l} \right) } } \right]^{\mathsf T},$
	where $\eta=\tfrac{2\pi u }{\lambda}$, $u$ is the antenna spacing, and $\lambda$ is the channel wavelength (Note that we also define $\upsilon_{l}=\eta \sin \left (\phi _{l} \right) $). Equivalently, the channel vector in ~(\ref{channel_model}) can be expressed in matrix-vector form as
	\begin{align}
	{\bf h}_{}= \sqrt { \frac{1}{ L} }{\bf A}_{} {\bf B}_{} {\bm \alpha}_{},
	\end{align}
	where
	\begin{align}
	&{\bf A}_{N\times L} = [{\bf a}\left (\phi _{1} \right), \dots, {\bf a}\left (\phi _{L} \right)]\label{Channel_parm},\notag\\ &{\bf B}_{L\times L} = \mathsf{diag}(\sqrt{\beta_{1}},\dots, \sqrt{\beta_{L}}), \text{ and } \notag\\ & {\bm \alpha}_{L\times 1} = [\alpha_{1},\dots,\alpha_{L}]^{\mathsf{T}}.
	\end{align}
	
	As mentioned previously, the quantities ${\bm \alpha}_{}$ are dependent on frequency; however ${\bf B}_{}$ and ${\bf A}_{}$ are constant with respect to frequency over an \textit{angle-coherence} time interval (as discussed in subsection ~\ref{sub:AngleCohTime}).
	
	To model a realistic system where we have non-ideal angle reciprocity, we assume that the differences between uplink and downlink multipath components, $\tilde{\upsilon}^{\mathrm{u/d}}_l$ and $\tilde{\beta}^{\mathrm{u/d}}_l$, are i.i.d. random variables with zero mean and variance $\sigma^{2}_\upsilon$, $\sigma^{2}_\beta \ll 1$ \cite{hugl2002spatial}.

	\subsection{Uplink  Training}
	{	Let ${\bf p}_{k} \in \mathbb{C}^{1\times \tau}$ be the uplink (UL) pilot signal sent by the $\nth{k}$ user composed of $\tau$ symbols with unit norm. All pilot sequences used by different users are assumed to be pairwise orthogonal, since the angle coherence time is much longer than the conventional channel coherence time~\cite{Heath2017Beamwidth}. Therefore, we can assign a sufficiently large number to $\tau$ such that $\tau\geq K$ holds true.}
	
	Therefore, the received signal ${\bf Y}_{mk} \in \mathbb{C}^{N\times \tau}$ at the $\nth{m}$ AP sent by the $\nth{k}$ user is given by
	\begin{equation}\label{rec_pilot}
	{\bf Y}_{mk}=\sqrt{\rho} {\mathbf{h}}_{mk} {{\bf p}_{k}}+ {\bf N}_{mk},
	\end{equation}
	where $\rho$ is the uplink transmit power and  {the entries $[{\bf N}_{mk}]_{n,i}$ of the additive white Gaussian noise matrix ${\bf N}_{mk} \in \mathbb{C}^{N\times \tau}$  are independent and identically distributed (i.i.d.) $\mathcal{CN}(0,\sigma_{n}^{2})$ random variables.} Multiplying ~(\ref{rec_pilot}) by ${\bf p}_{k}^\mathsf{H}$ and collecting $T$ samples, we have
	\begin{align}\label{rec_pilot2}
	{\bf Y}_{mk}(t){\bf p}_{k}^\mathsf{H}
	&=\sqrt{\frac{\rho}{L}} {\bf A}_{mk} {\bf B}_{mk} {\bm \alpha}_{mk}(t) + {\bf N}_{mk}{\bf p}_{k}^\mathsf{H}\notag\\ &=\sqrt{\rho} {\bf A}_{mk} {\bf d}_{mk}(t) + \bar{{\bf n}}_{mk}, \:\:\:t=1,\dots,T,
	\end{align}
	where ${\bf d}_{mk}=\tfrac{1}{\sqrt L}{\bf B}_{mk} {\bm \alpha}_{mk}$ and {$\bar{{\bf n}}_{mk}={\bf N}_{mk}{\bf p}_{k}^\mathsf{H} \sim  \mathcal{CN}({\bf 0}_{N\times1 },\sigma_n^2\mathbf{I}_N)$}. Then, the $T$  samples of ~(\ref{rec_pilot2}) are collected in a matrix form as
	\begin{align}\label{rec_pilot_BIG}
	{\bf \bar{Y}}_{mk}
	&=\sqrt{\rho} {\bf H}_{mk}  + {\bf \bar{ N}}_{mk}=\sqrt{\rho} {\bf A}_{mk} {\bf D}_{mk}  + {\bf \bar{ N}}_{mk},
	\end{align}	
	where ${\bf \bar{Y}}_{mk}=[{\bf Y}_{mk}(1){\bf p}_{k}^\mathsf{H}, \dots, {\bf Y}_{mk}(T){\bf p}_{k}^\mathsf{H}] $, ${\bf H}_{mk} = [{\bf h}_{mk}(1), \dots, {\bf h}_{mk}(T)]$,
	${\bf D}_{mk} = [{\bf d}_{mk}(1), \dots, {\bf d}_{mk}(T)]$, and ${\bf \bar{ N}}_{mk}= [\bar{{\bf n}}_{mk}(1),\dots, \bar{{\bf n}}_{mk}(T)]$.

	The multipath components estimation is performed in a distributed fashion, in which each AP \textit{independently} estimates the multipath components to the $K$ users. The APs do not cooperate on the multipath components estimation, and no estimates need to be shared among the APs.
	\subsection{Downlink Payload Data Transmission}
	
	The APs, based on the estimated multipath components, independently apply ${N\times 1}$ beamforming vector $\hat{\bf w}_{mk} $ to transmit signals to the $K$ users. Moreover, APs do not cooperate on the beamforming vectors.
	The transmit DL signal from the $\nth{m}$ AP is given by
	\begin{equation}\label{trans_APm}
	{\bf x}_{m}=\sqrt{\rho^{\mathrm d}}\sum_{k=1}^{K} \hat{\bf w}_{mk} s_{k}^{\mathrm d},
	\end{equation}
	where $s_{k}^{\mathrm d}$ is the data symbol for the $\nth{k}$ user satisfying $\mathbb{E}[|s_{k}^{\mathrm d}|^{2}]=1$, and ${\rho^{\mathrm d}}$ is the maximum transmit power satisfying, $	\mathbb{E}[||{\bf x}_{m}||^{2}_2] \leq {\rho^{\mathrm d}}$. {It can be noted here that the multiplexing order is equal to 1. }
	
	Then, the received downlink signal at the $\nth{k}$ user is given by
	\begin{align}
	r_{k}^\mathrm {d}		&=\sum_{m=1}^{M} {\bf h}_{mk}^\mathsf{H} {\bf x}_{m}+n_{k}^{\mathrm d}\notag\\ &=\underbrace{\sqrt{\rho^{\mathrm d}} \sum_{m=1}^{M} {\bf h}_{mk}^\mathsf{H}  \hat{\bf w}_{mk} s_{k}^{\mathrm d}}_{S}+\underbrace{\sqrt{\rho^{\mathrm d}} \sum_{j\neq k}^{K}\sum_{m=1}^{M} {\bf h}_{mk}^\mathsf{H}  \hat{\bf w}_{mj} s_{j}^{\mathrm d}}_{I}+n_{k}^{\mathrm d},
	\end{align}
	where $n_{k}^{\mathrm d} \sim \mathcal{CN}(0,1)$ is the additive noise at the $\nth{k}$  user. Note that the received signal can be decomposed into three parts: 1) desired signal part ($S$), 2) interference part ($I$), and 3) noise $n_{k}^{\mathrm d}$. Moreover, the $\nth{k}$ user can detect signal $s_{k}^{\mathrm {d}}																																																																																																																		$ from $r_{k}^{\mathrm {d}}$.

	\subsection{Uplink Payload Data Transmission}
	
	In the uplink, all $K$ users simultaneously send their data symbols  $s_k^{\mathrm u}$, where $ {\mathrm {\mathbb {E}}}\left \{{{|s_{k}^{\mathrm u}|^{2}}}\right \}=1$, to the APs.  {It can be noted here that the multiplexing order is equal to 1. } The received UL signal at the $\nth{m}$ AP is given by
	\begin{equation}
	{\bf y}_{m}^{\mathrm u}= \sqrt { \rho^{\mathrm {u}}} \sum _{k=1}^{K} {\bf h}_{mk} s_{k}^{\mathrm u}+ {\bf n}_{m}^{\mathrm u},
	\end{equation}
	where ${ \rho^{\mathrm {u}}}$ is the  uplink transmit power and ${\bf n}_{m}^{\mathrm u}$ is additive noise at the $\nth{m}$ AP. {The noise entries ($[{\bf n}_{ m}^{\mathrm u}]_i$) are modeled as i.i.d.  $\mathcal{CN}(0,\sigma_n^2)$}. The received signal is multiplied by the ${N\times 1}$ combiner $\hat{\bf v}_{mk} $ at each AP where the resulting signal is sent to the CPU through a backhaul to detect the signal. The CPU will receive
	\begin{align}
	r_{k}^{\mathrm u}=\sum _{m=1}^{M}\hat {\bf v}_{mk}^\mathsf{H} {\bf y}_{m}^{\mathrm u}=\sum _{k'=1}^{K}\sum _{m=1}^{M} \sqrt { \rho^{\mathrm {u}}} \hat {\bf v}_{mk}^\mathsf{H} {\bf h}_{mk'} s_{k'}^{\mathrm u} + \sum _{m=1}^{M} \hat {\bf v}_{mk}^\mathsf{H} {\bf n}_{m}^{\mathrm u}.\qquad
	\end{align}
	Then, $s_k$ is detected from $r_{k}^{\mathrm u}$.
	
	The main system parameters are summarized in Table~\ref{table:parameters}.
	\begin{table}[t!]
		\caption{System Parameters }
		\footnotesize
		\centering
		\label{table:parameters}
		\begin{tabular}{l|l}\hline
			Number of APs, and number of antennas per AP& $M, N$ \\
			Total number of users & $K$ \\
			Number of paths & $L$ \\
			Channel gain for the $\nth{m}$ AP and $\nth{k}$ user & ${\bf h}_{mk}$ \\
			Angular steering vector for the $\nth{l}$ path & ${\bf a}(\phi_l)$ \\
			Angular steering matrix for the $\nth{m}$ AP and $\nth{k}$ user  & ${\bf A}_{mk}$ \\
			Large scale fading matrix  & ${\bf B}_{mk}$ \\
			Small scale fading vector & ${\bm \alpha}_{mk}$ \\
			$N \times N$ DFT matrix& ${\bf F}_{N}$ \\\hline
		\end{tabular}
	\end{table}
	
	\section{Proposed Angle information aided channel estimation for FDD systems}\label{s:angle_estimation}
	In this section, we present the FDD-based cell-free massive MIMO systems that directly acquire multipath components from the uplink pilot signal and use them for the AP cooperation. Using array signal processing, we first present the low complexity DFT-based AoA estimation, and then we propose the large-scale fading estimation based on the estimated angle information. Note that we need to estimate both components (AoA, and large-scale fading) for every angle coherence interval, in order to apply low complexity beamforming/combining techniques.

	\subsection{AoA Estimation Algorithm}
	{Based on our previous work \cite{Asmaa2019cellfree}, we apply AoA estimation step that relies on the classical DFT estimation and angle rotation. DFT is used to estimate the AoA wherein the peak of the DFT magnitude spectrum can \textit{select} the column whose steering angle best matches the true AoA.}
	
	Moreover, the normalized DFT of the channel matrix is defined as $ {{\mathbf{h}}_{mk}^{\text{DFT}}}={\bf F}_{N}{\bf h}_{mk}$ where ${\bf F}_{N}$ is an $N \times N$ DFT matrix whose $\nth{(n,q)}$ element is given by $[{\bf F}_{N}]_{nq} = \tfrac{1}{\sqrt{N}} e^{-j\tfrac {2 \pi nq}{N}}$. Most of the channel power is concentrated around  $L$ largest peaks determined by the $(\nth{\lfloor q_l \rceil})$ elements where $ q_l =\tfrac{N \upsilon_{l,mk} }{2\pi}$ (for $l=1, \cdots, L$) and $\upsilon_{l,mk}=\eta\sin \left (\phi _{l,mk}\right)$ \cite{GAO2018AODest}. Therefore, the initial AoA estimate for the $\nth{k}$ user is $	\hat{\phi}_{l,mk}^{\text{ini}}=\sin^{-1}\left( \frac{\lambda q_l^\text{ini}}{Nd}\right)$.
	
	{Furthermore, the accuracy of the AoA estimation is improved through an angle rotation operation~\cite{GAO2018AODest} by incorporating a phase-shift to the initial estimation to obtain more accurate peaks.} The angle rotation of the original channel matrix is expressed as $	{\mathbf{h}}_{l,mk}^{\text{rDFT}}={\bf \Phi}_N(\triangle\phi_{l,mk}){{\mathbf{h}}_{mk}^{\text{DFT}}}$,  where ${\bf\Phi}(\triangle\phi_{l,mk})\!=\!\mathrm{diag}\left\{\!\left[\text{1},e^{j\triangle\phi_{l,mk}},\ldots,e^{j(N-1)\triangle\phi_{l,mk}}\right]\!\right\}$
	%
	with $\triangle \phi_{l}\in[-(\pi/N),\pi/N]$ is the angle rotation parameter.
	It is shown in \cite{GAO2018AODest}  that the entries of  ${[{{\mathbf{h}}_{mk}^\text{rDFT}}]}$ have only $L$ non-zero peak elements when the optimal phase shifter satisfies $\triangle \phi_{l,mk}= 2\pi q_{l}/N-\upsilon_{l,mk}=2\pi q_{l}/N-\eta \sin \left (\phi _{l,mk}\right)$.
	
	Therefore, the estimate $\hat{\phi}_{l,mk}$  can be expressed as $	\hat{\phi}_{l,mk}=\sin^{-1}\left( \frac{2\pi q_l}{N\eta} -\frac{ \triangle\phi_{l,mk}}{\eta}\right)$,
	and the estimated AoA matrix is given by
	\begin{align}\label{estimated_AOA}
	\hat{\bf A}_{mk} = \left[{\bf a}\left (\hat{\phi}_{1,mk} \right), \dots, {\bf a}\left (\hat{\phi}_{L,mk} \right)\right].
	\end{align}

	\begin{algorithm*}\small
		\caption{Extended DFT and Angle-Rotation-Based Multipath Component Estimation}\label{Alg:Ch_Est}
		\begin{algorithmic}[1]
			\State \textbf{Input:} ${\bf \bar{Y}} \in \mathbb{C}^{N\times T}$, $L$, $\mathcal{G}$ and $\lambda$
			\State\textbf{Output:} $\hat{\bm{\phi}} \in \mathbb{R}^{L\times 1}$, $\hat{\bm{\beta}}\in \mathbb{C}^{L\times 1}$
			\State {//} AoA Estimation
			\For {$\:\:l=1: L$}
			\For  {$\:\:t=1: T$}
			\State Find the central point ($q^{\text{ini}}_l$) of each bin in ${\hat{\mathbf{h}}}_{mk}^{\text{DFT}}={\bf F}_{N}{\bf \bar{y}}_{mk}^{\mathrm p}(t)$ where $(q^{\text{ini}}_l)= \arg \max_{(q)\in \text{bin}(l)}\Vert [{\hat{\mathbf{h}}}_{mk}^{\text{DFT}}]_q \Vert^2,\:\: l=1,\cdots\: L.$
			\State   $(\hat{\triangle \phi_{l}})=\arg \max_{\triangle \phi \in \mathcal{G}} \Vert {\bf f}_{Nq^{\text{ini}}_l}{\bf\Phi}(\triangle \phi_{l}){\bf \bar{y}}_{mk}^{\mathrm p}(t)\Vert^2,$ where ${\bf f}_{Nq^{\text{ini}}_l}$ is the $\nth{q^{\text{ini}}_l}$ column of ${\bf F}_{N}$.
			\State      $\hat{\theta}_{l}(t)=\hat{\theta}_{l}(t-1)+\sin^{-1}\left( \frac{2\pi q^{\text{ini}}_l}{N\eta} -\frac{ \triangle\phi_{l}}{\eta}\right)$
			\EndFor
			\State \textbf{end}
			\State $\hat{\phi}_{l,mk}=\frac{1}{T}\hat{\theta}_{l}(T)$
			\EndFor
			
			\State \textbf{end}
			\State {//} Large scale fading Estimation
			\State $\hat{\bf D}_{mk}= \frac{1}{\sqrt{\rho}}\left(\hat{\bf A}_{mk}^\mathsf{H}\hat{\bf A}_{mk}\right)^{-1}\hat{\bf A}_{mk}^\mathsf{H} {\bf \bar{Y}}_{mk}$, where $\hat{\bf A}_{mk} = [{\bf a}\left (\hat{\phi}_{1,mk} \right), \dots, {\bf a}\left (\hat{\phi}_{L,mk} \right)]$
			\State $\hat{\bf R}_{d} = \tfrac{L}{T}[\hat{\bf D}_{mk}\hat{\bf D}_{mk}^\mathsf{H}]$
			\State $\hat{\bm{\beta}}_{mk}=[\hat{\beta}_{1,mk},\dots,\hat{\beta}_{L,mk}]^{\mathsf{T}}=\mathsf{diag}(\hat{\bf R}_{d})$
			\State \textbf{end}
		\end{algorithmic}
	\end{algorithm*}

	\subsection{Large-Scale Fading Estimation}
	{Based on the AoA estimate and given that $\bar{{\bf n}}_{mk}\! \sim \!  \mathcal{CN}({\bf 0}_{N\times1 },\sigma_n^2\mathbf{I}_N)$ in~\eqref{rec_pilot_BIG}, the probability density function of ${\bf \bar{Y}}_{mk}$ for given $\phi_{l,mk}$ and $\beta_{l,mk}$ over all $l\!=\!1, \cdots, L$ can be expressed as
		\begin{align}\label{equ:pdf_y}
		f({\bf \bar{Y}}_{mk}|\phi_{l,mk},\beta_{l,mk})=\frac{\exp{\{-\tfrac{1}{\sigma_n^2}||{\bf \bar{Y}}_{mk}\!-\sqrt{\rho} {\bf A}_{mk} {\bf D}_{mk} ||_{\text{F}}^2\}}}{({\pi\sigma_n^2})^{N}}.
		\end{align}
		The log-likelihood function can be applied to~\eqref{equ:pdf_y} to give
		\begin{align}\label{log_likelihood}\small
		\!\!\mathcal{L}({\bf D}_{mk},\sigma_n^2)\!=\!-{N}\ln \pi-{N}\ln \sigma_n^2-\! \tfrac{||{\bf \bar{Y}}_{mk}\!-\sqrt{\rho} {\bf A}_{mk} {\bf D}_{mk} ||_{\text{F}}^2}{\sigma_n^2}.
		\end{align}}
	Knowing that $\mathcal{L}$ is a concave function of $\sigma_n^2$ and ${\bf D}_{mk}$, the optimal estimates $\hat{\sigma_n}^2$ and $\hat{\bf D}_{mk}$ can be obtained by taking a partial derivative with respect to $\sigma_n^2$ and ${\bf D}_{mk}$. Hence, $\hat{\sigma}_{n}^2= \tfrac{1}{N}||{\bf \bar{Y}}_{mk}-\sqrt{\rho} \hat{\bf A}_{mk} \hat{\bf D}_{mk} ||_{\text{F}}^2,$
	and
	\begin{align}\label{D_estimated}
	&\hat{\bf D}_{mk}= \tfrac{1}{\sqrt{\rho}}\hat{\bf A}_{mk}^\dagger{\bf \bar{Y}}_{mk},
	\end{align}
	%
	%
	%
	where $\hat{\bf A}_{mk} = [{\bf a} (\hat{\phi}_{1,mk} ), \dots, {\bf a} (\hat{\phi}_{L,mk} )]$ is the estimate of ${\bf A}_{mk}$ which is obtained using array signal processing (DFT operation with angle rotation). Once $\hat{\bf A}_{mk}$ is obtained, we next estimate the large-scale fading coefficients
	$\beta_{l,mk}$. From ~(\ref{D_estimated}), we can estimate ${\bf D}_{mk}$ and the covariance matrix $\hat{\bf R}_{mk,d} = \tfrac{L}{T}\mathbb{E}[\hat{\bf D}_{mk}\hat{\bf D}_{mk}^\mathsf{H}]$. Note that the original covariance
	matrix ${\bf R}_{mk,d}$ is given by
	\begin{align}
	{\bf R}_{mk,d} &=L \times \mathbb{E}[{{\bf d}_{mk}}{{\bf d}_{mk}}^\mathsf{H}]={\bf B}_{mk}\mathbb{E}[{{\bm \alpha}_{mk}}{{\bm \alpha}_{mk}}^\mathsf{H}]{\bf B}_{mk}^\mathsf{H}\notag\\ &= \mathsf{diag}(\beta_{1,mk},\dots,\beta_{L,mk}).
	\end{align}
	
	Hence, we can obtain the estimates of the large-scale fading coefficients as
	\begin{equation}\label{estimated_beta}
	\hat{\bm{\beta}}_{mk}=[\hat{\beta}_{1,mk},\dots,\hat{\beta}_{L,mk}]^T=\mathsf{diag}(\hat{\bf R}_{mk,d}).
	\end{equation}
	
	{The proposed multipath component estimation is shown in Algorithm \ref{Alg:Ch_Est}, where $\mathcal{G}$ is the search grid within $[-\tfrac{\pi}{N},\tfrac{\pi}{N}]$ needed for angle estimation.
		
		Note that the search grid parameter $\mathcal{G}$ determines the complexity and accuracy of the algorithm. The complexity of the whole algorithm is of the order ${O}(N\log N + \mathcal{G}NL)$ where the factor $N\log N$ comes from the DFT operation and $\mathcal{G}NL$ comes from rotation operation over a search grid $\mathcal{G}$ for all paths $L$ over $N$ antennas.  Moreover, the complexity of the proposed algorithm is less than that of the classical subspace ESPRIT algorithm of complexity $O(N^3+UN^2)$, with $U\gg \mathcal{G}$ being the number of snapshots required during blind estimation \cite{Stoeckle2015espiritCom}.}

	\subsection{Performance Analysis}
	
	{Using the same methodology as in \cite{Gao2017DoAestMmwave,GAO2018AODest} in addition to estimating the large-scale fading parameter, we derive the theoretical MSE of the AoA estimates and the large-scale fading coefficients for the cell-free massive MIMO system. In general, a closed-form solution of the MSE for multiple AoA estimations is hard to obtain~\cite{Gao2017DoAestMmwave}. An alternative approach is to consider the single user and single propagation path and derive corresponding MSE of $\phi$ and $\beta$ as benchmark~\cite{Gao2017DoAestMmwave}.}
	
	For a single propagation path according to (\ref{rec_pilot_BIG}), the received training signal at the $\nth{m}$ AP transmitted by the $\nth{k}$ user is given by
	\begin{align}\small
	\bar{{\bf y}}_{mk}&={{\bf Y}}_{mk}{\bf p}_{k}^\mathsf{H}=\sqrt{\rho} {\mathbf{h}}_{mk}  + \bar{\bf n}_{mk}=\sqrt{\rho} {\mathbf{a}}(\phi) d_{mk}+ \bar{\bf n}_{mk}\notag\\ &=\sqrt{\rho} \sqrt{ \beta_{mk}} \alpha_{mk}{\mathbf{a}}(\phi) + \bar{\bf n}_{mk},
	\end{align}
	where ${\mathbf{a}}(\phi)$ is the $N\times 1$ steering vector with its $\nth{q}$ entry given by $[{\mathbf{a}}(\phi)]_q=\tfrac{1}{\sqrt{N}}  e^{(q-1)\upsilon_{mk}}$.
	
	For brevity, we henceforth omit the subscript $mk$ representing the link between the $\nth{m}$ AP and the $\nth{k}$ user. The proposed angle estimator can be expressed as
	\begin{align}\label{prop_estimate}
	\hat{\upsilon}&= \arg \max\limits _{\upsilon} \Vert \tfrac{1}{\Vert{{\mathbf{{a}}(\phi)}}\Vert^2}{{\mathbf{{a}}(\phi)}}^\mathsf{H}{ \bf \bar{y}}_{} \Vert^2 = \arg \max\limits _{\upsilon} \Vert {{\mathbf{{a}}(\phi)}}^\mathsf{H}{ \bf \bar{y}}_{} \Vert^2\notag\\ & = \arg \max\limits _{\upsilon}{ \bf \bar{y}}_{}^\mathsf{H} {{\mathbf{{a}}(\phi)}}{{\mathbf{{a}}(\phi)}}^\mathsf{H} { \bf \bar{y}}_{},
	\end{align}
	where ${{\mathbf{{a}}(\phi)}}= {\bf \Phi}(\triangle \phi){\bf f}_{N_q} \text{, }{\Vert{{\mathbf{{a}}(\phi)}}\Vert^2}=1$, ${\bf f}_{Nq^{}}$ is the $\nth{q^{}}$ column of ${\bf F}_{N}$, and $q$ is the nearest integer to $\tfrac{N \upsilon}{2\pi}$.
	
	{Moreover, using ~(\ref{D_estimated}), the ML estimate of $d$ is obtained as}
	\begin{align}
	\hat{d}_{\mathsf{ML}}&=\tfrac{1}{\sqrt{\rho^{}} }({\mathbf{{a}}(\hat\phi)}^\mathsf{H}{\mathbf{{a}}(\hat\phi)})^{-1}{\mathbf{{a}}(\hat\phi)}^\mathsf{H} \bar{\bf y}
	=\tfrac{1}{\sqrt{\rho^{}}\Vert{{\mathbf{{a}}(\hat\phi)}}\Vert^2}{{\mathbf{{a}}(\hat\phi)}}^\mathsf{H} \bar{\bf y}\notag\\ &= \tfrac{1}{\sqrt{\rho^{}}\Vert{\mathbf{{a}}(\hat\phi)}\Vert^2}{\mathbf{{a}}(\hat\phi)}^\mathsf{H} {\mathbf{{a}}(\phi)}d+\tfrac{1}{\sqrt{\rho^{}}\Vert{\mathbf{{a}}(\hat\phi)}\Vert^2}{\mathbf{{a}}(\hat\phi)}^\mathsf{H}\bar{\bf n}\notag\\
	&=\tfrac{1}{\sqrt{\rho^{}}\Vert{\mathbf{{a}}(\hat\phi)}\Vert^2}{\mathbf{{a}}(\hat\phi)}^\mathsf{H} {\mathbf{{a}}(\phi)}\sqrt{\beta}\alpha+\tfrac{1}{\sqrt{\rho^{}}\Vert{\mathbf{{a}}(\hat\phi)}\Vert^2}{\mathbf{{a}}(\hat\phi)}^\mathsf{H}\bar{\bf n}.\label{d_estimate}
	\end{align}

	{	The joint ML estimates of $\upsilon$ and $d$ can be obtained from}
	\begin{equation}\label{ML_estimate}
	[\hat{\upsilon}_{\mathsf{ML}} \hat{d}_{\mathsf{ML}}]= \arg \min\limits _{\upsilon, d} \Vert \bar{\bf y}_{}- {\mathbf{a}}(\phi)d\Vert^2,
	\end{equation}
	where $\hat{\upsilon}_{\mathsf{ML}}, \hat{d}_{\mathsf{ML}}$ are the optimizing variables.

	{	Therefore, using~\eqref{d_estimate}, the ML estimate of $\upsilon$ is given by}
	\begin{align}
	\hat{\upsilon}_{\mathsf{ML}}=\arg \max\limits _{\upsilon}\bar{\bf y}^\mathsf{H} {\bf P_a}\bar{\bf y} =\arg \max\limits _{\upsilon} g(\upsilon),
	\end{align}
	where $g(\upsilon)$ is the cost function of $\upsilon$. For the single-path case, $\bf {P_a}\!=\!\bf{a}(\phi) \bf{a}(\phi)^\mathsf{H}$ is the projection matrix onto the subspace spanned by $\bf{a}(\phi)$, and $\bf{a}(\phi)$ represents the steering vector given in ~\eqref{channel_model}. For the multi-path case, $\bf {P_A}\!=\! \bf{A} \bf{A}^\dagger \!=\!\bf{A} (\bf{A}^\mathsf{H}\bf{A})^{-1}\bf{A}^\mathsf{H}$ represents the projection matrix onto the subspace spanned by $\bf A$, and $\bf A$ is the steering matrix given in~\eqref{Channel_parm}. As shown in~\cite{GAO2018AODest}  while including the large scale path-loss parameter $\beta$, the MSE~\eqref{prop_estimate} of the considered DFT estimator coincides with that of the ML estimator~\eqref{ML_estimate}. Using Lemma~1 in~\cite{GAO2018AODest} while including the large-scale fading parameter and ${{\bf p}_k{\bf p}_k^\mathsf{H}}\!=\!1$, the MSE of $\upsilon$ is expressed as
	\begin{equation}\label{equ:theo_MSE}
	\expec{\triangle \upsilon^{2}}=\mathbb{E}[(\hat{\upsilon}-\upsilon)(\hat{\upsilon}-\upsilon)^{\mathsf{H}}] =\frac{\sigma_{n}^{2}}{2{\rho^{}}  {\beta} \mathbf{a}(\hat \phi)^\mathsf{H}\mathbf{E}\mathbf{P_{a}^{\perp}}\mathbf{E}\mathbf{a}(\hat\phi)},
	\end{equation}
	where  $\expec{\triangle \upsilon}=0$, $\mathbf{P_{a}^{\perp}}= \mathbf{I}-\mathbf{P_{a}}$ is the projection matrix onto the orthogonal space spanned by $\bf a(\phi)$ and $\mathbf{E}$ is the diagonal matrix given by $	\mathbf{E}=\mathsf{diag}\{{0,\cdots,(N-1)}\}.$
	%
	Based on the fact that $\upsilon= \eta\sin\phi$ and $\phi= \sin^{-1} (\tfrac{ \upsilon}{\eta})$, we further examine the MSE of $\phi$
	\begin{equation}
	\expec{\triangle \phi^{2}}=\frac{(\tfrac{1}{\eta})^2}{1-(\tfrac{\upsilon}{\eta})^2}\times\frac{\sigma_{n}^{2}}{2 \beta \mathbf{a}(\hat\phi)^\mathsf{H}\mathbf{E}\mathbf{P_{a}}^{\perp}\mathbf{E}\mathbf{a}(\hat\phi)}.
	\end{equation}
	
	Using Taylor series expansion, a of first-order approximation of ${\bf{a}(\phi)}$ is given by
	\begin{equation}\label{TaylorsA}
	{\bf{a}(\phi)}={\bf{a}}(\hat\phi)+j\mathbf{E}{\bf{a}}(\hat\phi)\triangle \upsilon.
	\end{equation}
	
	Substituting ~(\ref{TaylorsA}) into ~(\ref{d_estimate}) and after collecting $T$ samples, we rewrite $\hat{\bf d}$ as
	\begin{align}
	\hat{\bf d} &=[\hat{d}_1,\cdots,\hat{d}_T]\notag\\ &= {\bf d}+j\tfrac{1}{\Vert{\mathbf{{a}}(\hat\phi)}\Vert^2}{{\bf{a}}(\hat\phi)}^\mathsf{H}\mathbf{E} {\bf{a}}(\hat\phi)\triangle \upsilon {\bf d}+\tfrac{1}{\sqrt{\rho^{}}\Vert{\mathbf{{a}}(\hat\phi)}\Vert^2}{{\bf{a}}(\hat\phi)}^\mathsf{H}\bar{\bf N},
	\end{align}
	where $\bar{\bf N}=[\bar{\mathbf{n}}_1,\cdots,\bar{\mathbf{n}}_T]$.
	
	Moreover,
	\begin{align}
	\hat{{\beta}}&= {L\over T}\mathbb{E}[{\bf \hat{d}}{\bf \hat{d}}^\mathsf{H}]=\beta+ \beta \expec{(\triangle \upsilon)^2} \vert \tfrac{1}{\Vert{\mathbf{{a}}(\hat\phi)}\Vert^2}{{\bf{a}}(\hat\phi)}^\mathsf{H}\mathbf{E} {\bf{a}}(\hat\phi)\vert^2 +\notag\\ &~~ \tfrac{1}{\sqrt{\rho^{}}\Vert{\mathbf{{a}}(\hat\phi)}\Vert^2}{{\bf{a}}(\hat\phi)}^\mathsf{H}\expec{{\bf nn}^\mathsf{H}}(\tfrac{1}{\sqrt{\rho^{}}\Vert{\mathbf{{a}}(\hat\phi)}\Vert^2}{\mathbf{{a}}}(\hat\phi)^\mathsf{H})^\mathsf{H} \notag\\
	&={\beta}+\frac{\sigma_{n}^{2}{\vert\mathbf{{a}}(\hat\phi)}^\mathsf{H}\mathbf{E} {\mathbf{{a}}(\hat\phi)}\vert^2}{2\rho^{} {\mathbf{{a}}(\hat\phi)}^\mathsf{H}\mathbf{E}\mathbf{P_{a}}^{\perp}\mathbf{E}{\mathbf{{a}}(\hat\phi)}}+\frac{\sigma_{n}^2}{\rho^{}}.
	\end{align}
	
	Therefore, the MSE of $\beta$ can be obtained
	\begin{align}\label{MSE_beta}
	\expec{\triangle \beta^2}&=\expec{(\hat{\beta}-\beta)(\hat{\beta}-\beta)^\mathsf{H}}\notag\\ &= \left(\frac{\sigma_{n}^{2}\vert {\mathbf{{a}}(\hat\phi)}^\mathsf{H}\mathbf{E} \mathbf{ {a}}(\hat\phi)\vert^2}{2\rho \mathbf{{a}}(\hat\phi)^\mathsf{H}\mathbf{E}\mathbf{P_{a}}^{\perp}\mathbf{E}\mathbf{{a}}(\hat\phi)}+\frac{\sigma_{n}^2}{\rho^{}}\right)^2.
	\end{align}
	
	Furthermore, the MSE expressions of the estimated AoA and large-scale fading components derived in~\eqref{equ:theo_MSE} and~\eqref{MSE_beta} give important insights when assessing the impact of beamforming/combining techniques on the spectral efficiency of the proposed FDD-based cell-free massive MIMO system.
	
	\subsection{Angle Coherence Time}\label{sub:AngleCohTime}
	Different from the conventional channel coherence time, the angle coherence time is defined as typically an order of magnitude longer, during which the AoDs can be regarded as static \cite{Heath2017Beamwidth}. Specifically, the path AoD in ~(\ref{channel_model}) mainly depends on the surrounding obstacles around the BS, which may not physically change their positions often. On the contrary, the path gain of the $\nth{k}$ user depends on a number of unresolvable paths, each of which is generated by a scatter surrounding the user. Therefore, path gains vary much faster than the path AoDs \cite{Heath2017Beamwidth}. Accordingly, the angle coherence time is much longer than the conventional channel coherence time. Therefore, we can leverage from this fact and  perform multipath estimation in every angle coherence time instead of the much shorter channel coherence time as the impact of the overhead is substantially reduced.

	\section{Proposed Beamforming and Combining Techniques}\label{s:Beam_Com}
	{We next propose the angle-based matched-filtering, angle-based  zero-forcing and angle-based minimum-mean-square-error  beamforming/combining that incorporate the estimated angle information, and the large-scale fading components.}
	
	The APs are connected via a backhaul network to a CPU, which sends to the APs the data-symbols to be transmitted to the end-users and receives soft-estimates of the received data-symbols from all the APs. Neither multipath estimates nor beamforming/combining vectors are transmitted through the backhaul network.
	
	\subsection{Angle-Based Beamforming} \label{Sec:Beamforming}
	The angle-based beamforming (or precoding) vector $\hat{\bf w}_{mk}$ for the $\nth{m}$ AP and the $\nth{k}$ user is defined as
	\begin{equation}\label{beamforming}
	\hat{\bf w}_{mk}= \sum_{l=1}^{L} \gamma_{mk,l} \hat{\bf g}_{mk,l}=\frac{\hat{\bf G}_{mk}}{||\hat{\bf G}_{mk}||}{\bm \gamma}_{mk},
	\end{equation}
	where $\hat{\bf g}_{mk,l}$ is the $\nth{l}$ column of $\hat{\bf G}_{mk}=[\hat{\bf g}_{mk,1},\dots,\hat{\bf g}_{mk,L}]$ defined below for the proposed angle-based beamforming techniques. In addition, $\gamma_{mk,l}$ is the normalized complex weight for the $\nth{l}$ propagation path that satisfies $\sum_{l=1}^{L}|\gamma_{mk,l}|^{2}=1$ and ${\bm \gamma}_{mk}=[\gamma_{mk,1},\dots,\gamma_{mk,L}]^T$. Moreover, using ~(\ref{trans_APm}),
	\begin{equation}\label{equ:power_cons}
	\mathbb{E}[||{\bf x}_{m}||^{2}]={\rho_d}\sum_{k=1}^{K}\tfrac{||\hat{\bf G}_{mk}{\bm \gamma}_{mk}||^2}{||\hat{\bf G}_{mk}||^2} \leq {\rho_d}
	\end{equation}
	will satisfy the maximum transmit power $\rho_d$.
	\subsubsection{Angle-Based Matched-Filtering Beamforming (A-MF)}
	
	The precoder matrix based on the angle information is given by
	\begin{equation}\label{equ:MF}
	\hat{\bf G}_{mk}^{\textrm{A-MF}}=\hat{\bf A}_{mk}\hat{\bf B}_{mk},
	\end{equation}
	where $\hat{\bf A}_{mk}=\left[{\bf a}\left (\hat\phi _{1,mk} \right), \dots, {\bf a}\left (\hat\phi _{L,mk} \right)\right]$  and $\hat{\bf B}_{mk} = \mathsf{diag}\left(\sqrt{\hat\beta_{1,mk}},\dots, \sqrt{\hat\beta_{L,mk}}\right)$ are the estimated AoA and large-scale fading matrices according to ~(\ref{estimated_AOA}) and ~(\ref{estimated_beta}). {Moreover, A-MF is a simple beamforming approach that only requires the channel multipath components (AoA and large-scale fading) of the direct link between the $\nth{m}$ AP and the $\nth{k}$ user. However, the inter user interference is ignored.}
	\subsubsection{Angle-Based Zero-Forcing Beamforming (A-ZF)}
	
	We use A-ZF beamforming as a means to efficiently suppress interference. To do so, the conventional ZF beamforming employs all the downlink CSI from the users. However, the angle-based ZF beamforming used in this work is distinct from the conventional ZF beamforming in the sense that only the angle information and large-scale fading coefficients of the channel are required in the beamforming design.
	We collect the corresponding array steering vectors into
	$\hat{\bf A}_{m}=[\hat{\bf A}_{m1},\dots, \hat{\bf A}_{mK}]$ and similarly for $\hat{\bf B}_{m}=\mathsf{diag} \left([\hat{\bf B}_{m1},\dots, \hat{\bf B}_{mK}]^\mathsf{T}\right)$.  Then, the precoder matrix is given by
	\begin{equation}\label{equ:ZF}
	\hat{\bf G}_{m}^{\textrm{A-ZF}}=\hat{\bf A}_{m}\hat{\bf B}_{m}\left(\hat{\bf B}_{m}^\mathsf{H} \hat{\bf A}_{m}^\mathsf{H}\hat{\bf A}_{m}\hat{\bf B}_{m}\right)^{-1},
	\end{equation}
	where beamforming vector is $\hat{\bf g}_{mk,l}$ defined as the $\nth{((k-1)L+l)}$ column of $\hat{\bf G}_{m}^{\textrm{A-ZF}}$.
	
	A key property of the angle-based ZF beamforming is that the beamforming vector is orthogonal to all other array steering vectors as given below:
	\begin{equation}\label{equ:vec_detec}
	\hat{ \mathbf {h}}_{mk}^\mathsf{H}\hat{\bf w}_{mi}^{\textrm{A-ZF}} = \left\{
	\begin{array}{c l}
	{\bf s}_{mk}^T {\bm \gamma}_{mk}  & \text{if}\: i=k;\\
	0 & \text{if}\: i\neq k .
	\end{array}
	\right.
	\end{equation}
	
	The pseudo-inverse in A-ZF is more complex than A-MF, but the interference is suppressed.

	\subsubsection{Angle-Based MMSE Beamforming (A-MMSE)}{
		We use an angle-based MMSE beamforming design that can efficiently suppress interference, noise and channel estimation error. The A-MMSE strikes a balance between attaining the best signal amplification and reducing the interference. The proposed angle-based MMSE beamforming matrix is given by
		
		\begin{align}\label{equ:MMSE}\small
		&{\mathbf{G}}_{mk}^{\textrm{A-MMSE}}=\notag\\ &\left(\sum_{k=1}^{K}((\hat{\bf A}_{mk}\hat{\bf B}_{mk}\hat{\bf B}_{mk}^\mathsf{H} \hat{\bf A}_{mk}^\mathsf{H}+ \Upsilon_{m,k})+ \sigma_n^2 {\bf I}_{N}\right)^{-1}\hat{\bf A}_{mk}\hat{\bf B}_{mk},
		\end{align}
		\normalsize
		where $\Upsilon_{m,k}=\tilde{\sigma}^2_{\upsilon}(\mathbf{E}\hat{\bf{A}}_{mk}\hat{\bf B}_{mk})(\mathbf{E}\hat{\bf{A}}_{mk}\hat{\bf B}_{mk})^\mathsf{H}+\tilde{\sigma}^2_{\upsilon}\tilde{\sigma}^2_{\beta}(\mathbf{E}\hat{\bf{A}}_{mk})(\mathbf{E}\hat{\bf{A}}_{mk})^\mathsf{H}+\tilde{\sigma}^2_{\beta}{\hat{\bf{A}}}_{mk}{\hat{\bf{A}}}_{mk}^\mathsf{H},$
	
		such that  $\tilde{\sigma}^2_{\upsilon}=\sigma^{2}_{\upsilon}+\expec{\triangle \upsilon^2}$ and  $\tilde{\sigma}^2_{\beta}=\sigma^{2}_{\beta}+\expec{\triangle \beta^2}$, where $\sigma^{2}_{\upsilon}$ and $\sigma^{2}_{\beta}$ account for non-ideal DL angle reciprocity, and $\expec{\triangle \upsilon^2}$, $\expec{\triangle \beta^2}$ are the MSEs as defined in~(\ref{equ:theo_MSE}) and ~(\ref{MSE_beta}), respectively.

		Therefore, for A-ZF/A-MMSE, the only overhead for DL channel acquisition at each AP comes from UL training, which only scales with the number of served users. {In addition, one can note that A-ZF is suitable for high signal-to-noise ratio (SNR) conditions since it is expected that A-ZF and A-MMSE would have the same performance when the effect of noise is low.}

		\subsection{Angle-Based Combining}\label{Sec:Combining}
		Similarly, the combining  vector $\hat{\bf v}_{mk}$ for the $\nth{m}$ AP and the $\nth{k}$ user is defined as
		\begin{equation}\label{combining}
		\hat{\bf v}_{mk}= \sum_{l=1}^{L} \gamma_{mk,l} \hat{\bf c}_{mk,l}={\hat{\bf C}_{mk}}{\bm \gamma}_{mk},
		\end{equation}
		where $\hat{\bf c}_{mk,l}$ is the $\nth{((k-1)L+l)}$ column of $\hat{\bf C}_{m}$ which corresponds to $\hat{\bf C}_{mk}=[\hat{\bf c}_{mk,1},\dots,\hat{\bf c}_{mk,L}]$, and $\gamma_{mk,l}=\tfrac{1}{L}$ and ${\bm \gamma}_{mk}=[\gamma_{mk,1},\dots,\gamma_{mk,L}]^T$.
		
		Using UL-DL duality \cite{bjornson2017massive}, the combining vectors of the uplink case for A-MF combining, A-ZF combining and A-MMSE combining are also defined as
		\begin{equation}\label{equ:v}
		\hat{\bf C}_{mk}=\left\{
		\begin{array}{c l}
		{\mathbf{G}}_{mk}^{\textrm{A-MF}} & \text{for A-MF combining};\\
		{\mathbf{G}}_{mk}^{\textrm{A-ZF}} & \text{for A-ZF combining};\\
		{\mathbf{G}}_{mk}^{\textrm{A-MMSE}} & \text{for A-MMSE combining}.
		\end{array}
		\right.
		\end{equation}
		such that $\tilde{\sigma}^2_{\upsilon}=\expec{\triangle \upsilon^2}$ and  $\tilde{\sigma}^2_{\beta}=\expec{\triangle \beta^2}$.
		The corresponding combining matrices were defined in ~(\ref{equ:MF}), ~(\ref{equ:ZF}) and ~(\ref{equ:MMSE}).
		
		The benefits of relying on only the angle information and large-scale fading are: (i) the need for downlink training is avoided; (ii) the beamforming/combining matrices can be updated every angle coherence time, and (iii) a simple closed-form expression for the spectral efficiency can be derived which enables us to obtain important insights.
		
		\section{Spectral and Energy efficiency Analysis}\label{s:Spectral_efficiency_Analysis}
		In this section, we derive closed-form expressions for the {spectral efficiencies} per user for DL and UL transmissions using the analysis technique from~\cite{Ngo2017CellFreevsSmallCell,Ngo2018EEofCF,Kim2018CFreeFDD}. Then, we define the total energy efficiency of the system.
		
		\subsection{Spectral Efficiency}
		
		The downlink {spectral efficiency} per user using the proposed beamforming schemes is given by
		\begin{align}\label{equ:Rdsim}
		&R^{\mathrm d}_k	= {\log_2 \left( 1+ \mathrm{SINR_k^d} \right)}\simeq{\log_2 \left( 1+ \frac{{\rho^{\mathrm d}}S_k^{\mathrm d}}{{{\rho^{{\mathrm d}}}I_{jk}^{\mathrm d}}+{\rho^{{\mathrm d}}}{BU_k^{\mathrm d}}+\sigma_{n}^{2}} \right)},
		\end{align}
		where
		\begin{align}
		&S_k^{\mathrm d}=\sum_{m=1}^{M}\expec{||\hat{\bf h}_{mk}^\mathsf{H}  \hat{\bf w}_{mk}||^{2}},\notag\\ & I_{jk}^{\mathrm d}=\sum_{j\neq k}^{K}\sum_{m=1}^{M}\expec{||\hat{\bf h}_{mk}^\mathsf{H}  \hat{\bf w}_{mj}||^{2}}, \text{ and }\notag\\ &BU_k^{\mathrm d}=\sum_{j= 1}^{K}\sum_{m=1}^{M}\expec{||\tilde{\bf h}_{mk}^\mathsf{H}  \hat{\bf w}_{mj}||^{2}},\notag
		\end{align}
		represent the strength of the desired signal of the $\nth{k}$ user ($S_k^{\mathrm d}$), the interference generated by the $\nth{j}$ user ($I_{jk}^{\mathrm d}$), and the beamforming gain uncertainty ($BU_k^{\mathrm d}$), respectively. The elements inside the norm of $S_k^{\mathrm d}$, $I_{jk}^{\mathrm d}$ and $BU_k^{\mathrm d}$  are uncorrelated zero mean random variables. In addition, $\hat{\bf h}_{mk}={\bf h}_{mk}-\tilde{\bf h}_{mk}=\hat{\bf A}_{mk} \hat{\bf B}_{mk} {\bf s}_{mk}$ and the channel uncertainty is $\tilde{\bf h}_{mk}=\triangle \tilde{\upsilon} ({\bf{E}}\hat{\bf A}_{mk}\hat{\mathbf{B}}_{mk}){\bf s}_{mk}+\triangle \tilde{\beta} \hat{\bf A}_{mk}{\bf s}_{mk}+\triangle \tilde{\beta}  \triangle \tilde{\upsilon} {\bf E}\hat{\mathbf{A}}_{mk}{\bf s}_{mk}$, where $\triangle \tilde{\upsilon} $ and $\triangle \tilde{\beta}$ differ in the DL and UL directions due to un-ideal angle reciprocity such that $\triangle \tilde{\upsilon} ^{\mathrm{d}}= \upsilon^{\mathrm{u}}-\hat{\upsilon}^{\mathrm{u}}-\tilde{\upsilon}^{\mathrm{u/d}}$,  $\triangle \tilde{\beta} ^{\mathrm{d}}= \beta^{\mathrm{u}}-\hat{\beta}^{\mathrm{u}}-\tilde{\beta}^{\mathrm{u/d}}$,  $\triangle \tilde{\upsilon} ^{\mathrm{u}}= \upsilon^{\mathrm{u}}-\hat{\upsilon}^{\mathrm{u}}$, and $\triangle \tilde{\beta} ^{\mathrm{u}}= \beta^{\mathrm{u}}-\hat{\beta}^{\mathrm{u}}$.
		
		Similarly for the uplink case, the uplink {spectral efficiency} per user using the proposed combining schemes is given by
		\begin{align}\label{equ:Rusim}
		&R^{\mathrm u}_k\simeq{\log_2 \left( 1+ \frac{{\rho^{\mathrm u}}S_k^{\mathrm u}}{{{\rho^{\mathrm u}}I_{jk}^{\mathrm u}}+{\rho^{\mathrm u}}{BU_k^{\mathrm u}}+\sigma_{n}^{2}\sum_{m=1}^{M}||\hat{\bf v}_{mk}||^{2}} \right)},
		\end{align}
		where uplink desired signal power ($S_k^u$), the interference caused by the $\nth{j}$ user ($I_{jk}^u$), and the combining gain uncertainty ($BU_k^u$) are defined similarly as the  downlink case but by substituting $\hat{\bf w}_{mj}$ with the combining vector $\hat{\bf v}_{mj}$.
		
		Using the fact that $\alpha_l \sim \mathcal{CN}(0,1)$ as well as the fact that angle of arrival and the large-scale fading remain unchanged during the angle coherence time, we can further reduce the DL and UL spectral efficiencies into closed forms as shown in ~(\ref{equ:Rd1}) and ~(\ref{equ:Ru1}) at the top of the next page.
		\begin{figure*}\label{equ:Rd}
			\begin{align}\label{equ:Rd1}
			R^{\mathrm d}_k	&\simeq{}\log_2 \left( 1+ \frac{{\rho^{\mathrm d}}\sum_{m=1}^{M}{||{\bf \hat B}_{mk}^\mathsf{H} {\bf \hat A}_{mk}^\mathsf{H} \hat{\bf w}_{mk}||^{2}}}
			{{\rho^{\mathrm d}}\sum_{j\neq k}^{K}\sum_{m=1}^{M}{||{\bf \hat B}_{mk}^\mathsf{H} {\bf \hat A}_{mk}^\mathsf{H}  \hat{\bf w}_{mj}||^{2}}+{\rho^{\mathrm d}}\sum_{j=1}^{K}\sum_{m=1}^{M}\Omega_{m,j}+\sigma_{n}^{2}} \right),
			\end{align}
			
			where $\Omega_{m,j}={ \tilde{\sigma}^2_{\upsilon}}\Vert(\hat{\bf{B}}_{mk}^\mathsf{H}\hat{\bf A}_{mk}^\mathsf{H}\mathbf{E})\hat{\bf w}_{mj}\Vert^2+ \tilde{\sigma}^2_{\beta}\Vert(\hat{\bf A}_{mk}^\mathsf{H}\hat{\bf w}_{mj}\Vert^2+ \tilde{\sigma}^2_{\beta} \tilde{\sigma}^2_{\upsilon}\Vert(\hat{\bf A}_{mk}^\mathsf{H}\mathbf{E}\hat{\bf w}_{mj}\Vert^2.$
		\end{figure*}
		\begin{figure*}\label{equ:Ru}
			\begin{align}\label{equ:Ru1}
			R^{\mathrm u}_k	\simeq{}\log_2 \left( 1+ \frac{{\rho^{\mathrm u}}\sum_{m=1}^{M}{||{\bf \hat B}_{mk}^\mathsf{H} {\bf \hat A}_{mk}^\mathsf{H} \hat{\bf v}_{mk}||^{2}}}
			{{\rho^{\mathrm u}}\sum_{j\neq k}^{K}\sum_{m=1}^{M}{||{\bf \hat B}_{mk}^\mathsf{H} {\bf \hat A}_{mk}^\mathsf{H}  \hat{\bf v}_{mj}||^{2}}+{\rho^{\mathrm u}}\sum_{j=1}^{K}\sum_{m=1}^{M}\Lambda_{m,j}+\sigma_{n}^{2}\sum_{m=1}^{M}||\hat{\bf v}_{mk}||^{2}} \right),
			\end{align}
			
			where $	\Lambda_{m,j}= \tilde{\sigma}^2_{\upsilon}\Vert(\hat{\bf{B}}_{mk}^\mathsf{H}\hat{\bf A}_{mk}^\mathsf{H}\mathbf{E})\hat{\bf v}_{mj}\Vert^2+\tilde{\sigma}^2_{\beta}\Vert(\hat{\bf A}_{mk}^\mathsf{H}\hat{\bf v}_{mj}\Vert^2+ \tilde{\sigma}^2_{\beta} \tilde{\sigma}^2_{\upsilon}\Vert(\hat{\bf A}_{mk}^\mathsf{H}\mathbf{E}\hat{\bf v}_{mj}\Vert^2.$
		\end{figure*}
		
		\vspace{-0.25in}
		\subsection{Energy Efficiency}
		The total energy efficiency (bit/Joule) is defined as the sum throughput (bit/s) divided by the total power consumption (Watt) in the network:
		\begin{equation} \label{EE}
		{\mathsf { EE}}\triangleq \frac {B{\cdot \sum_{k= 1}^{K} \kappa R_k }}{ P_{\mathrm {total}} },
		\end{equation}
		where $R_k$ is the spectral efficiency (expressed in bit/s/Hz) for the $\nth{k}$ user, $B$ is defined as the system bandwidth, $P_{\mathrm {total}}$ is the total power consumption, $\kappa=\left(1-\frac{\tau}{\tau_c}\right) $, and $\tau=K$ is length of pilot training sequence in samples, $\tau_c$ is the angle coherence interval in samples.
		Furthermore, we consider the power consumption model defined in \cite{Ngo2018EEofCF}
		\begin{equation}\label{equ:P_tot}
		P_{\mathrm {total}}= \sum _{m=1}^{M} P_{m} + \sum _{m=1}^{M} P_{\text {bh},m},
		\end{equation}
		where $P_{m} $ is the power consumed at the $\nth{m}$ AP which includes the amplifier and the circuit power consumption and the power consumption of the transceiver chains and the power consumed for signal processing, and $P_{\text {bh},m}$ represents the power consumed by the backhaul link that transfers data between the CPU and the $\nth{m}$ AP.  The power consumption term $P_m$ can be defined as
		\begin{equation}
		P_{m} = \frac {1}{\vartheta _{m}} \rho ^{\mathrm {d}} \sigma^2_{n} \left (N{ \sum _{k=1}^{K} ||\hat{\bf w} _{mk}||^2}\right ) + N P_{\text {tc},m},
		\end{equation}
		where $0<{\vartheta _{m}} \leq 1$ is the power amplifier efficiency, $\rho ^{\mathrm {d}}$ is the downlink SNR, $ \sigma^2_{n}$ is the noise power, $\hat{\bf w} _{mk}$ is the angle based beamforming vector for the $\nth{m}$ AP and the $\nth{k}$ user (defined in ~(\ref{beamforming})), $N$ is the number of antennas at the AP, and $P_{\text {tc},m}$ is the internal power required to operate the circuit components (e.g., converters, mixers, and filters) per antenna at the $\nth{m}$ AP.
		
		Moreover, the power consumption of the backhaul is proportional to the sum spectral efficiency and can be modeled as,\vspace{-1.5pt}
		\begin{equation} \label{equ:P_backhaul}
		P_{\text {bh},m} = P_{0,m} +B \cdot \sum_{k= 1}^{K} \kappa R_k \cdot P_{\text {bt},m},
		\end{equation}
		where $P_{0,m}$ is defined as a fixed power consumption of each backhaul (traffic-independent power) which may depend on the distances between the APs and the CPU and the system topology,  and $P_{\text {bt},m}$ is defined as the traffic-dependent power (in Watt per bit/s).

		\section{Proposed Max-Min Power control}
		To obtain good system performance, the available power resources must be efficiently managed. In this section, we propose a solution to the max-min user-fairness problem in the proposed cell-free Massive MIMO system, where the minimum uplink rates of all users are maximized while satisfying a per-user power constraint. We show that the FDD-based cell-free massive MIMO system can provide uniformly good service to all users, regardless of their geographical location, by adopting a max-min power/weight control strategy.
		The proposed power control algorithm is done at the CPU, and importantly, is carried only at the \textit{angle-coherence} time-scale. Hence the impact of the signaling overhead is substantially reduced. {Moreover, we present a user centric AP selection approach to further enhance the energy efficiency of the CF massive MIMO system.}
		
		\subsection{Downlink Power Control}\label{DL_PC}
		In the downlink, given realizations of the large-scale fading and the array steering vectors, we find the power control coefficients ${\bm \gamma}_{mk}$, $m = 1,\dots,M$, $k = 1,\dots,K,$ that maximize the minimum of the downlink rates of all users, under the power constraint ~(\ref{equ:power_cons}). At the optimum point, all users attain the same rate. Mathematically, this is formulated as:
		\begin{align}\label{equ:max_minPC1}
		&\max _{\{\gamma _{mk,l}\}}~\min \limits _{k=1, \cdots , K} R^{ \mathrm {d}}_k \notag
		\\&\text {subject to}~\sum_{k=1}^{K}\tfrac{||\hat{\bf G}_{mk}{\bm \gamma}_{mk}||^2}{||\hat{\bf G}_{mk}||^2} \leq 1, ~ m=1,\ldots , M \notag\\ &~~\qquad~ \gamma _{mk,l} \geq 0, ~ \forall k, ~ \forall m,  ~ \forall l.
		\end{align}
		
		Then, using ~(\ref{equ:Rd1}), we can reformulate ~(\ref{equ:max_minPC1}) into a max-min SINR problem as follows:
		\begin{align}\label{equ:power_control}\small
		&\max _{\{\gamma_{mk,l}\}}~\min \limits _{k=1, \cdots , K} \notag\\ &\tfrac{{\rho^{\mathrm d}}\sum_{m=1}^{M}{||{\bf \hat B}_{mk}^\mathsf{H} {\bf \hat A}_{mk}^\mathsf{H} \hat{\bf w}_{mk}||^{2}}} 	{{\rho^{\mathrm d}}\sum_{j\neq k}^{K}\sum_{m=1}^{M}{||{\bf \hat B}_{mk}^\mathsf{H} {\bf \hat A}_{mk}^\mathsf{H}  \hat{\bf w}_{mj}||^{2}}+{\rho^{\mathrm d}}\sum_{j=1}^{K}\sum_{m=1}^{M}\Omega_{m,j}+\sigma_{n}^{2}}\notag\\[0.5pt]
		&\text {s.t.}~\sum_{k=1}^{K}\tfrac{||\hat{\bf G}_{mk}{\bm \gamma}_{mk}||^2}{||\hat{\bf G}_{mk}||^2} \leq 1, ~ \forall m,
		\notag\\ &~~\hat{\bf w}_{mk}= \frac{\hat{\bf G}_{mk}}{||\hat{\bf G}_{mk}||}{\bm \gamma}_{mk},~ \forall k, ~ \forall m\text{, and }\notag\\ &~~\qquad~ \gamma _{mk,l} \geq 0, ~ \forall k, ~\forall m, ~\forall l.
		\end{align}
		One can note that ~(\ref{equ:power_control}) is a non-convex separable quadratically-constrained quadratic program (QCQP) in terms of power allocation ${\bm \gamma}_{mk}$, for all $k, m$. Therefore, this problem cannot be directly solved in an efficient manner using existing convex optimization schemes. While the non-convex QCQP is NP-hard, it can be relaxed into a convex semi-definite program (SDP) using semi-definite relaxation (SDR)~\cite{SDR2010Luo}, in which the following property of a scalar is utilized: $	{\bm \gamma}_{mk}^{\mathsf{H}}\mathbf{Q}{\bm \gamma}_{mk}= \text{tr} ({\bm \gamma}_{mk}^{\mathsf{H}}\mathbf{Q}{\bm \gamma}_{mk})= \text{tr} (\mathbf{Q}{\bm \gamma}_{mk}{\bm \gamma}_{mk}^\mathsf{H}),$
		for any $\mathbf{Q} \in \mathbb{C}^{L\times L}$. Therefore, by introducing a new variable ${\bf \Gamma}_{mk}={\bm \gamma}_{mk}{\bm \gamma}_{mk}^\mathsf{H}$, which is a rank-one symmetric positive semi-definite (PSD) matrix, the quadratic constraints can be transformed into linear constraints in the set of all real symmetric $L\times L$ matrices $\mathbb{S}^L$. Using SDP, problem ~(\ref{equ:power_control}) can be equivalently reformulated as
		\begin{align}\label{min_maxPC}\small
		&\max _{\{{\bf \Gamma}_{mk}\}}~\min \limits _{k=1, \cdots , K} \notag\\ &\tfrac{{\rho^{\mathrm d}}\sum_{m=1}^{M}{\text{tr}\left({\bf \Xi}_{mkk}{\bf \Xi}_{mkk}^{\mathsf{H}} {\bf \Gamma}_{mk}\right) }}	{{\rho^{\mathrm d}}\sum_{j\neq k}^{K}\sum_{m=1}^{M}{\text{tr}\left({\bf \Xi}_{mkj}{\bf \Xi}_{mkj}^{\mathsf{H}} {\bf \Gamma}_{mj}\right) }+{\rho^{\mathrm d}}\sum_{j=1}^{K}\sum_{m=1}^{M}\Omega_{m,j}+\sigma_{n}^{2}}\notag\\[0.5pt]
		&\text {s.t.}~\sum_{k=1}^{K}\tfrac{\text{tr}\left(\hat{\bf G}_{mk}^{\mathsf{H}}\hat{\bf G}_{mk}{\bf \Gamma}_{mk}\right)}{||\hat{\bf G}_{mk}||^2} \leq 1, ~~ \forall  m,\notag\\ &\qquad~{\bf \Gamma}_{mk}\succeq  0, ~  \forall k, ~ \forall m,\notag\\ &~\text{rank}\left({\bf \Gamma}_{mk}\right)=1 , ~  \forall k, ~ \forall m,
		\end{align}
		where ${\bf \Xi}_{mkj} ={\bf \hat B}_{mk}^\mathsf{H} {\bf \hat A}_{mk}^\mathsf{H} \tfrac{\hat{\bf G}_{mj}}{||\hat{\bf G}_{mj}||}$.\vspace{0.6pt}
		
		Since the rank constraint of ${\bf \Gamma}_{mk}$ is non-convex, we relax it to obtain the feasible SDP formulation of ~(\ref{min_maxPC}) as 
		\begin{align}\label{min_maxPCBS}\small
		&\max _{\{{\bf \Gamma}_{mk}\}} \mu \notag\\[0.5pt]
		&\text {s.t.}\notag \\&~ \tfrac{{\rho^{\mathrm d}}\sum_{m=1}^{M}{\text{tr}\left({\bf \Xi}_{mkk}{\bf \Xi}_{mkk}^{\mathsf{H}} {\bf \Gamma}_{mk}\right) }}	{{\rho^{\mathrm d}}\sum_{j\neq k}^{K}\sum_{m=1}^{M}{\text{tr}\left({\bf \Xi}_{mkj}{\bf \Xi}_{mkj}^{\mathsf{H}} {\bf \Gamma}_{mj}\right) }+{\rho^{\mathrm d}}\sum_{j=1}^{K}\sum_{m=1}^{M}\Omega_{m,j}+\sigma_{n}^{2}} \geq \mu,\notag\\[0.5pt]
		&~\sum_{k=1}^{K}\tfrac{\text{tr}\left(\hat{\bf G}_{mk}^{\mathsf{H}}\hat{\bf G}_{mk}{\bf \Gamma}_{mk}\right)}{||\hat{\bf G}_{mk}||^2} \leq 1, \forall  m, \text{and} \qquad~{\bf \Gamma}_{mk}\succeq  0, ~  \forall k, ~ \forall  m.
		\end{align}
		
		The relaxed problem ~(\ref{min_maxPCBS})  is a convex SDP and can be solved by standard convex optimization tools such as CVX~\cite{grant2008cvx}. Once the optimal variables $\hat {\bf \Gamma}_{mk}~ (\forall m, ~\forall k)$ are obtained, we can find the rank-one approximations of $\hat {\bf \Gamma}_{mk}$ which are feasible for the original problem ~(\ref{equ:power_control}) by applying eigen-value decomposition (EVD) on $\hat {\bf \Gamma}_{mk}$, and extracting the largest eigen-value and the corresponding eigen-vector to construct $\hat{\bm \gamma}_{mk}$. Consequently,~\eqref{min_maxPCBS} can be solved efficiently via a bisection search, in which each step involves solving a sequence of convex SDP feasibility subproblems~\cite{boyd2004convex}. The proposed max-min power control algorithm is summarized in Algorithm~\ref{Alg:PC}.
		
		{\textit{Complexity Analysis:} Here, we provide the computational complexity analysis for the proposed Algorithm~\ref{Alg:PC}, which uses iterative bisection search to solve the convex optimization problem ~(\ref{min_maxPCBS}) at each iteration. The complexity of ~(\ref{min_maxPCBS}) is $\mathsf{O}((MK)^{4}L^{1/2})$ in each iteration \cite{SDP1996Helm}. Note that the total number of iterations to solve the SDR Problem via a bisection search method is given by $\log (\tfrac{\mu_{\max}-\mu_{\min}}{\epsilon} )$,  where $\epsilon$ refers to a predetermined threshold \cite{boyd2004convex}. Hence, the total complexity of solving  ~(\ref{min_maxPCBS})  is $\mathsf{O}((MK)^{4}L^{1/2})\log (\tfrac{\mu_{\max}-\mu_{\min}}{\epsilon} )$.  }

		\begin{algorithm*}[t]\small
			\caption{SDR-based Bisection Algorithm for Solving~(\ref{min_maxPCBS})}\label{Alg:PC}
			\begin{algorithmic}[1]
				\State \textbf{Initialization:} Define the initial values $\mu_{\max}$, $\mu_{\min}$ that represent the range of relevant values of the objective function in~(\ref{min_maxPCBS}), and Choose a tolerance $\epsilon>0$
				\State\textbf{Set:} $\mu=\frac{\mu_{\max}+\mu_{\min}}{2}$,
				\State\textbf{Solve the following convex SDP feasibility program:}
				\State \begin{equation} \label{equ:min_maxoptima}\left\{
				\begin{array}{c }
				{{\rho^{\mathrm d}}\sum_{m=1}^{M}{\text{tr}\left({\bf \Xi}_{mkk}{\bf \Xi}_{mkk}^{\mathsf{H}} {\bf \Gamma}_{mk}\right) }}	 \geq \mu \left({{\rho^{\mathrm d}}\sum_{j\neq k}^{K}\sum_{m=1}^{M}{\text{tr}\left({\bf \Xi}_{mkj}{\bf \Xi}_{mkj}^{\mathsf{H}} {\bf \Gamma}_{mj}\right) }+{\rho^{\mathrm d}}\sum_{j=1}^{K}\sum_{m=1}^{M}\Omega_{m,j}+\sigma_{n}^{2}}\right), ~\forall  k, \\
				\sum_{k=1}^{K}\tfrac{\text{tr}\left(\hat{\bf G}_{mk}^{\mathsf{H}}\hat{\bf G}_{mk}{\bf \Gamma}_{mk}\right)}{||\hat{\bf G}_{mk}||^2} \leq 1,  ~\forall  m, \text{ and } \qquad~{\bf \Gamma}_{mk}\succeq  0, ~\forall  k, ~\forall  m,  \\
				\end{array}
				\right.
				\end{equation}
				\If{If problem ~(\ref{equ:min_maxoptima}) is feasible,}
				\State { set $\mu_{\min}=\mu$ }
				\Else { set $\mu_{\max}=\mu$.}\EndIf
				\State \textbf{end if}
				\State Stop \textbf{if }$\mu_{\max}-\mu_{\min}< \epsilon$. \textbf{Otherwise}, go to Step 2.
				\State $[{\bf U}_{mk},{\bf V}_{mk}]=\text{EVD}\left({\bf \Gamma}_{mk}\right), ~\forall  k, ~\forall  m, $ where ${\bf  V}_{L\times L}$ is the diagonal matrix of eigenvalues, and ${\bf U}_{L\times L}$ is a full matrix whose columns are the corresponding eigenvectors (${\bf u}$).
				\State ${\bm \gamma}_{mk}= \sqrt{\max ({\bf V}_{mk}) }{\bf u}_{m,k}^{\max}, ~\forall  k, ~\forall  m,$ where ${\bf u}^{\max} $ is the corresponding eigenvector to the maximum eigenvalue in ${\bf V}$.
				\State \textbf{end}
			\end{algorithmic}
		\end{algorithm*}
		
		\subsection{Uplink Weight Control}
		Similarly in the uplink, given realizations of the large-scale fading and the array steering vectors, we find the weight control coefficients ${\bm \gamma}_{mk}$, $m = 1,\dots,M$, $k = 1,\dots,K,$ that maximize the minimum of the uplink rates of all users, under the weight constraint. At the optimum point, all users attain the same rate. So,
		\begin{align}\label{equ:max_minULPC1}
		&\max _{\{\gamma _{mk,l}\}}~\min \limits _{k=1, \cdots , K} R^{ \mathrm {u}}_k \notag
		\\&\text {subject to}~\sum_{k=1}^{K}\tfrac{||\hat{\bf C}_{mk}{\bm \gamma}_{mk}||^2}{||\hat{\bf C}_{mk}||^2} \leq 1, ~ m=1,\ldots , M, \notag
		\\& \qquad~ \gamma _{mk,l} \geq 0, ~ \forall k, ~ \forall m,  ~ \forall l.
		\end{align}
		
		Moreover,~(\ref{equ:max_minULPC1}) can be solved following the same steps as shown in subsection (\ref{DL_PC}) in the DL case.
		
		\subsection{User-Centric (UC) AP Selection Method}
		{As noted from the last term in ~(\ref{equ:P_tot}) that represents the total power consumption of the backhaul, cell-free massive MIMO systems require more backhaul connections to transfer data between the APs and the CPU when compared to the co-located massive MIMO. Moreover, the second term of ~(\ref{equ:P_backhaul}) has a significant effect on the energy efficiency, especially when $M$ increases in ~(\ref{equ:P_tot}). To improve the total energy efficiency, we can further decrease the denominator of the energy efficiency in ~(\ref{EE}).
			We present an AP selection for the user-centric case which can reduce the backhaul power consumption, and hence, increase the energy efficiency. The AP selection scheme is based on choosing for each user $k$ a subset of APs $\mathcal{M}_k$ that forms ($\delta\%$) of the total channel power. For a particular user, there are many APs which are located very far away. These APs will not impact the overall spatial diversity gains. Hence, not all APs actually contribute in serving this user. Furthermore, $\mathcal{M}_k$ is chosen based on the following:
			\begin{equation}
			\sum_m^{\mathcal{M}_k} \frac{||{\bf A}_{mk}^{\star}{\bf B}_{mk}^\star||^2}{\sum_m^M{||{\bf A}_{mk}{\bf B}_{mk}||^2}}\geq \delta \%
			\end{equation}
			where  $\{||{\bf A}_{1k}^{\star}{\bf B}_{1k}^\star|| , \cdots , ||{\bf A}_{Mk}^{\star}{\bf B}_{Mk}^\star|| \}$ represents the sorted (in descending order) set of
			the set $\{||{\bf A}_{1k}{\bf B}_{1k}|| ,\cdots, ||{\bf A}_{Mk}{\bf B}_{Mk}||\}$. Therefore, by applying the presented AP selection scheme, each access point $m$ serves a subset $\mathcal{K}_m$ of $K$ users. Hence, the power allocation schemes proposed in the preceding subsections will allocate power ${\bm \gamma}_{mk}^\star={\bm \gamma}_{mk}$ if $k,m \in \mathcal{K}_m, \mathcal{M}_k$, respectively, and ${\bm \gamma}_{mk}^\star={\bf 0}_{L\times 1}$ otherwise. Therefore,  Algorithm \ref{Alg:PC}  can be directly applied where ${\bf \Gamma}_{mk}$ is replaced by ${\bf 0}_{L\times L}$ when $m \notin \mathcal{M}_k$ for $k \in \mathcal{K}_m$.
		}

		\section{Simulation Results}\label{sec:simulation}
		In this section, we study the performance of the proposed multipath components estimation compared to conventional schemes, and we provide numerical results to quantitatively study the performance of FDD cell-free massive MIMO in
		terms of downlink and uplink spectral efficiency for all the proposed beamforming and combining techniques.
		
		\subsection{Experimental Setup and Parameters}
		The APs and the users are located within a square of $1\times 1 \text{ km}^2$. The square is wrapped around at the edges to avoid boundary effects. Furthermore, for simplicity, random pilot assignment is used. With random pilot assignment, each user randomly chooses a pilot sequence from a predefined set of orthogonal pilot sequences of length $\tau =K$.
		The large-scale fading coefficient $\beta_{l,mk}$ is modeled as the product of path loss and shadow fading as in~\cite{Kim2018CFreeFDD}:
		\begin{align}\small
		&10\log_{10}(\beta_{l,mk})=\notag
		\\&\left\{
		\begin{array}{c l}
		P \unit[-37.6] \log_{10}(u_{mk})+z_{mk,l}- 15\log_{10}(u_{1}), & \!\!\!\!\text{if } u_{mk} > u_{1};\notag\\
		P \unit[-35] \log_{10}(u_{mk})+z_{mk,l}, &\!\!\!\! \text{if } u_{mk} \leq u_{1}.
		\end{array}
		\right.
		\end{align}
		\normalsize
		where $u_{mk}$ is the distance between the $\nth{m}$ AP and $\nth{k}$ user in kilometers, $z_{mk,l} \sim \mathcal{N}(0,\sigma_z^2)$  is the shadow fading variable with $\sigma_z=\unit[8]{dB}$, $u_{1}=\unit[0.05]{km}$ and $P=\unit[-148]{dB}$ for line-of-sight (LOS) and $P=\unit[-158]{dB}$ for non-line-of-sight (NLOS) propagation.
		
		{Moreover,  for the AP selection schemes, we choose $\delta=95\%$.} The system parameters used throughout the experimental simulations are summarized in Table \ref{table:1}.
		\small
		\begin{table}[t!]
			\centering
			\caption{Simulation Parameters}
			\footnotesize
			\label{table:1}
			\begin{tabular}{l|c}
				\hline
				\textbf{Parameter} & Value   \\\hline
				Cell radius ($D$) & $\unit[1]{km}$   \\\hline
				System Bandwidth ($B$) & {$\unit[100]{MHz}$}   \\\hline
				Uplink/Downlink Frequencies &  $\unit[49.8/50]{GHz}$    \\\hline
				Uplink pilot training transmit power $\rho$&  $\unit[200]{mW}$    \\\hline
				Uplink transmit power $\rho^{\mathrm u}$&  $\unit[200]{mW}$    \\\hline
				Downlink transmit power $\rho^{\mathrm d}$&  $\unit[1000]{mW}$    \\\hline
				Power amplifier parameter $\vartheta$ &  $0.2$    \\\hline
				Internal power consumption/each backhaul, $P_{\text{tc},m} \forall m$ \cite{Ngo2018EEofCF}&  $\unit[0.2]{W}$    \\\hline
				Fixed power consumption/each backhaul, $P_{0,m} \forall m$ \cite{Ngo2018EEofCF}&  $\unit[0.825]{W}$    \\\hline
				Traffic dependent  backhaul power, $P_{\text{bt},m} \forall m$ \cite{Ngo2018EEofCF}&  $\unit[0.25]{W/(Gbits/s)}$    \\\hline
				User Centric threshold ($\delta$) & $95\%$   \\\hline
				Angle coherence interval ($\tau_c$) & $200$ samples   \\\hline
				Monte-Carlo Simulations & 1000\\\hline
			\end{tabular}
		\end{table}
		\normalsize
		\begin{figure}[t]
			\centering \includegraphics[scale=0.34]{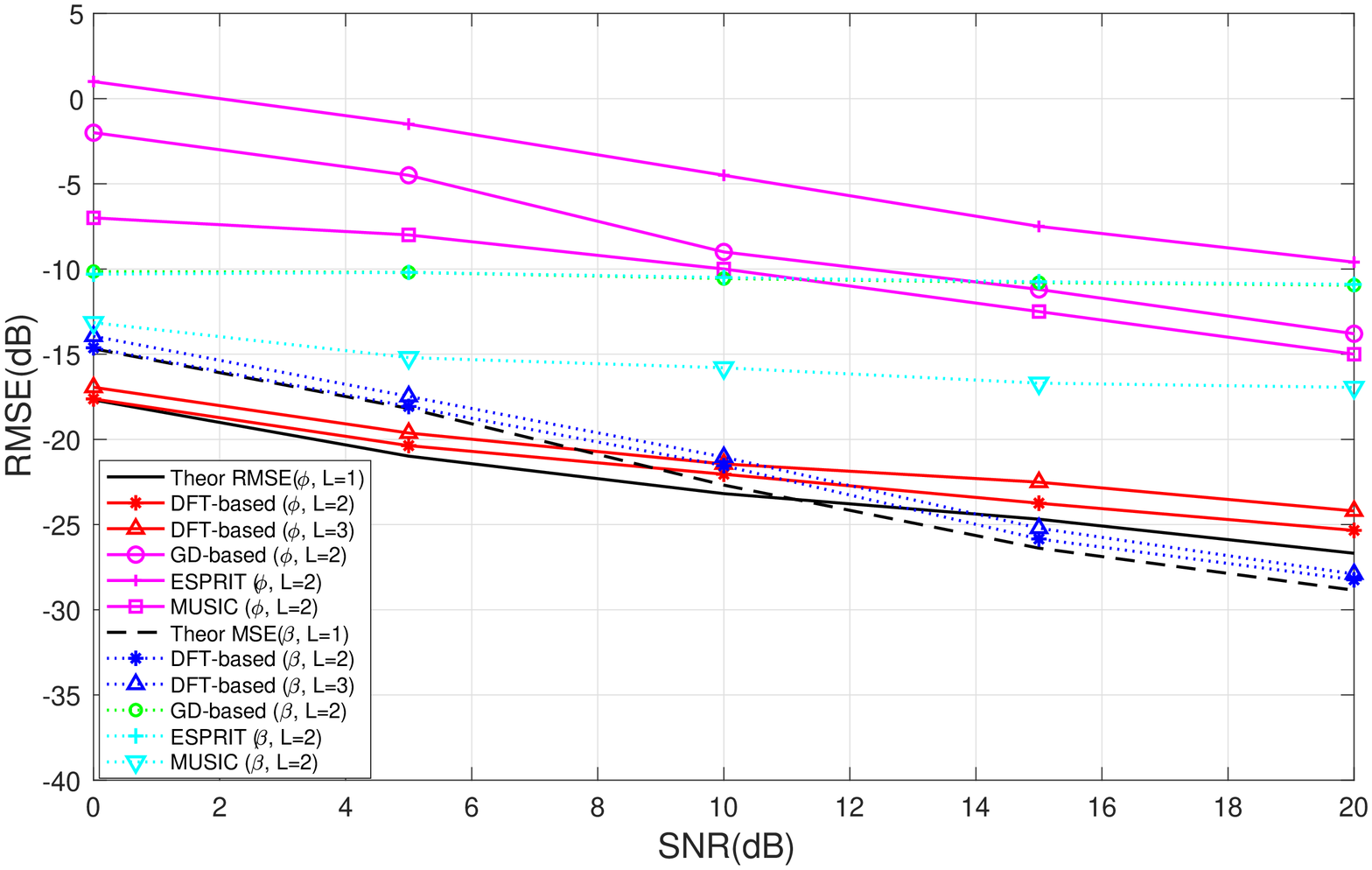}\vspace{-0.3in}
			\caption{RMSE performance of the multipath component estimation versus SNR for $N=32$ and $T=16$ compared with the gradient-descent based estimation and subspace-based estimation.}
			\label{fig:RMSE_theta_beta}
		\end{figure}

		\begin{figure*}[t!]
			\centering
			\subfigure[]{ \includegraphics[scale=0.36]{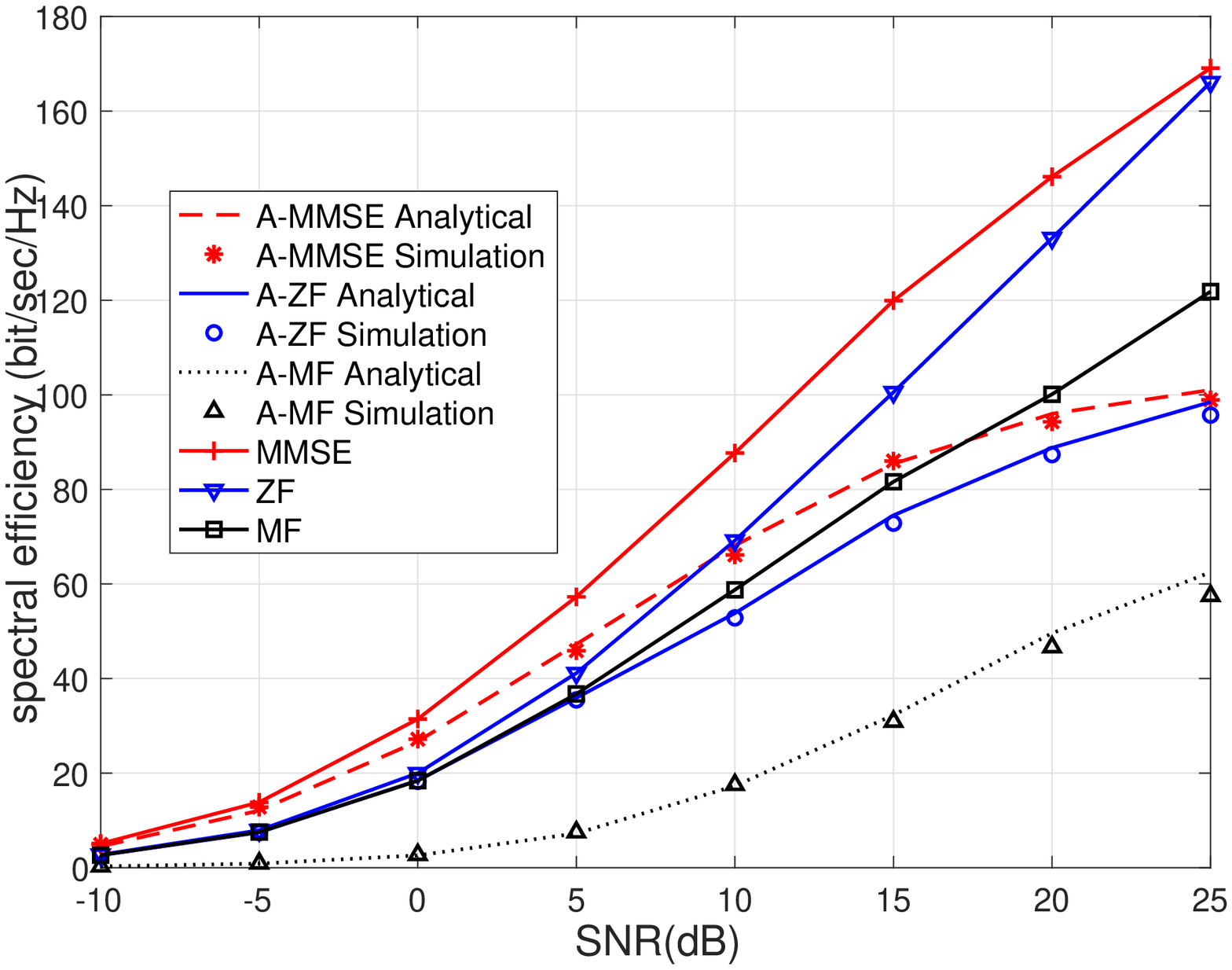}
				\label{fig:sum_Rate_DLimp}}\hspace{-0.35in}
			\subfigure[]{			\includegraphics[scale=0.36]{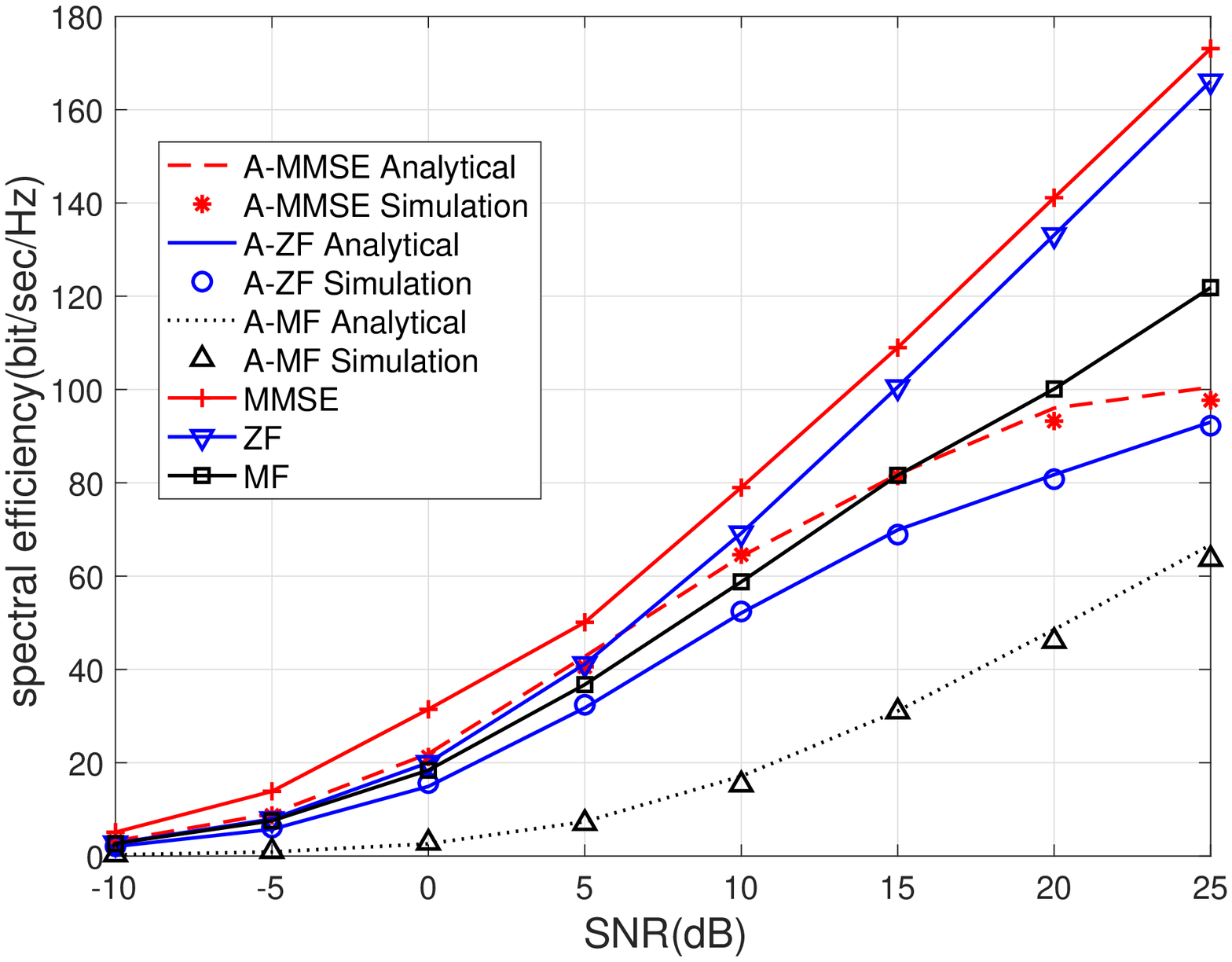}
				\label{fig:sum_Rate_ULimp}}\vspace{-0.15in}
			\caption{Spectral efficiency of the proposed beamforming schemes versus SNR for $M=10$ APs with $N=32$ antennas and $K=20$ users under imperfect channel estimation: (a) for DL, and (b) UL.}
			\label{fig:globfig2}
		\end{figure*}
		
		\begin{figure*}[t!]
			\centering
			\subfigure[]{ \includegraphics[scale=0.37]{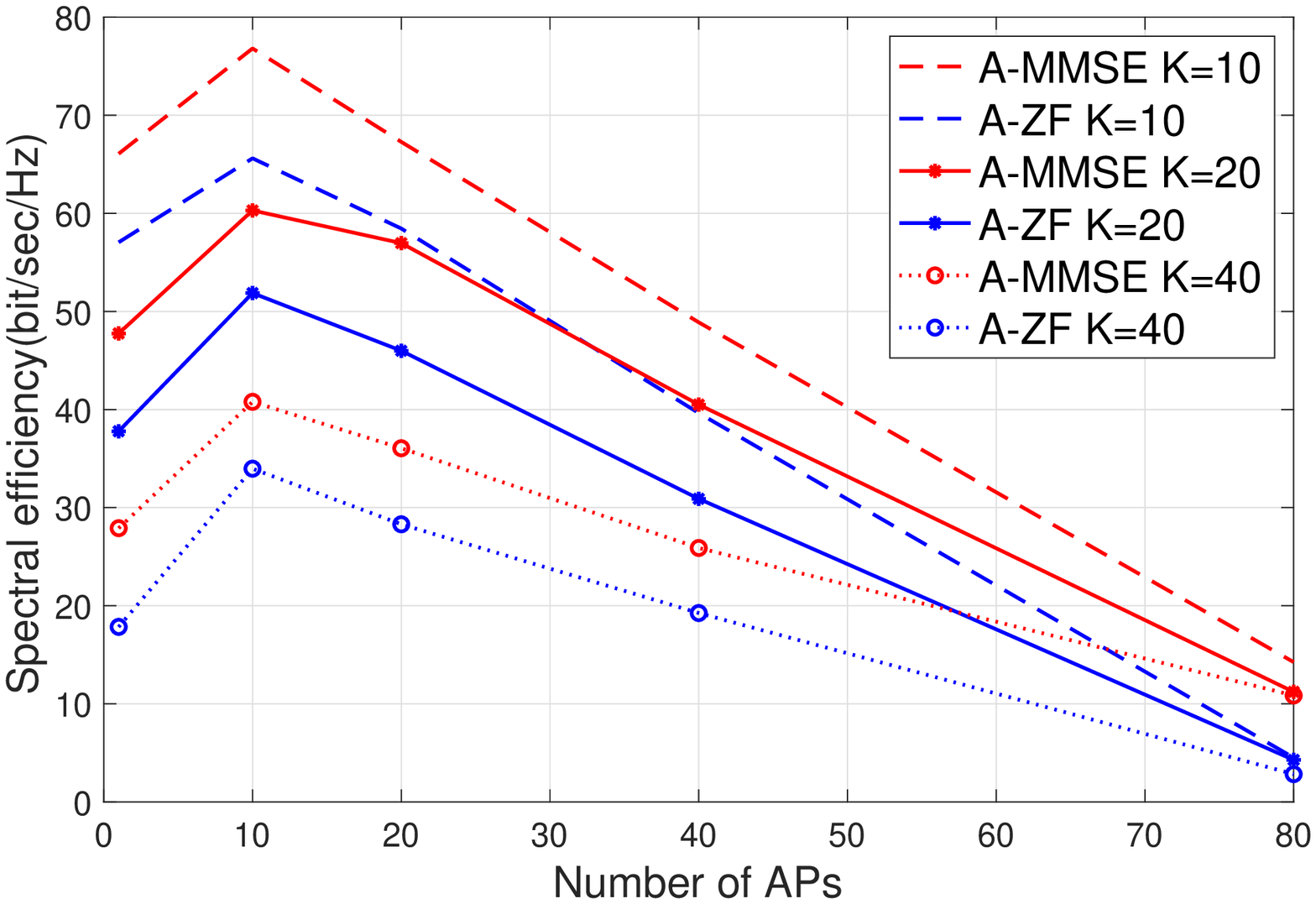}
				\label{fig:sum_Rate_variable_DL_NM}}
			\subfigure[]{		\includegraphics[scale=0.37]{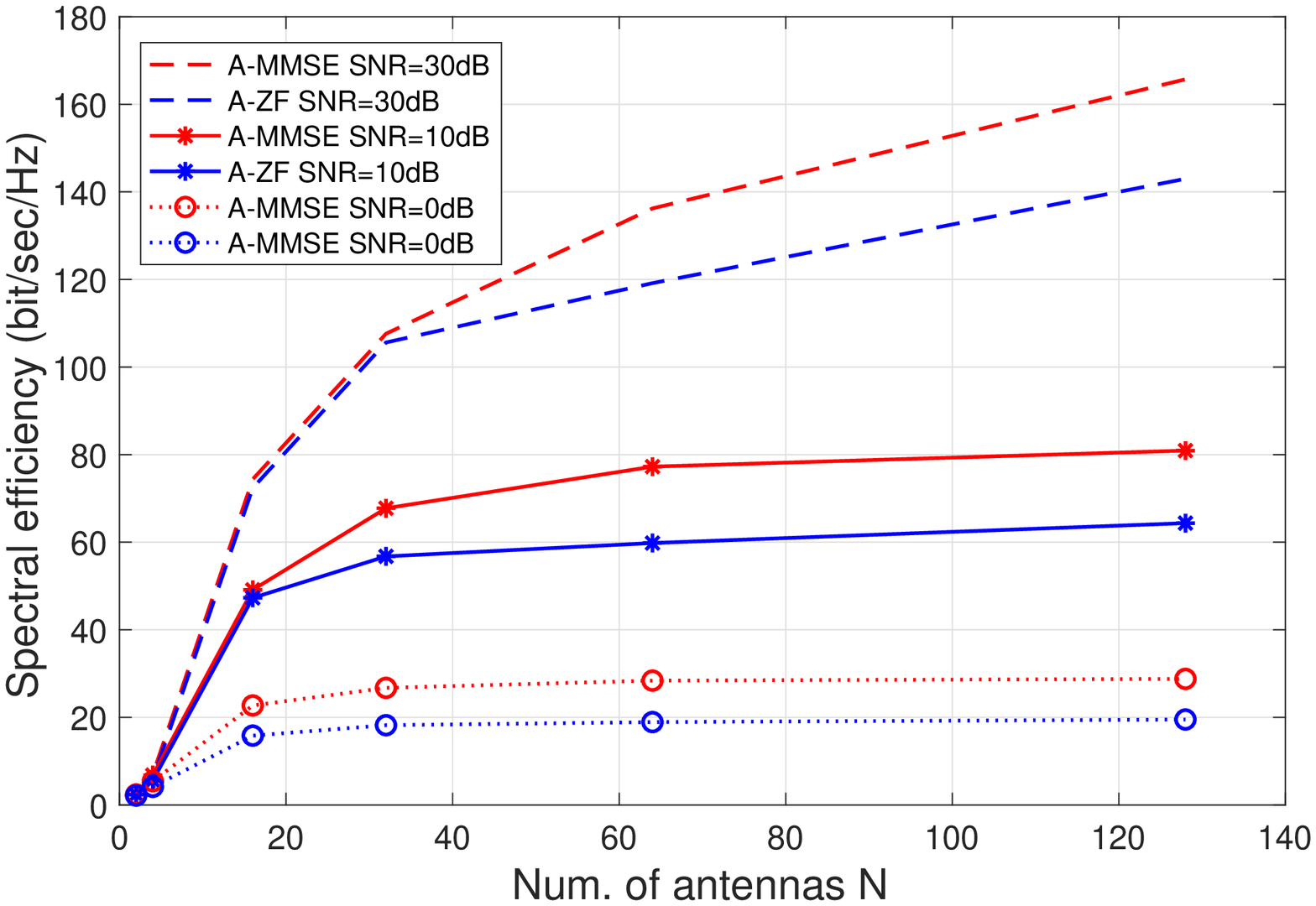}
				\label{fig:sum_Rate_variable_DL_M}}\vspace{-0.15in}
			\caption{DL sum-rate of the proposed combining schemes versus (a) number of APs at SNR=$\unit[10]{dB}$ for $MN=320$ and $K=\{10, 20, 40\}$ users, and (b) versus number of antennas $N$ at various SNR values for $M=10$ APs and $K=20$ users.}
			\label{fig:globfig4}
		\end{figure*}

		\begin{figure*}[t!]
			\centering
			\subfigure[]{
				\includegraphics[scale=0.37]{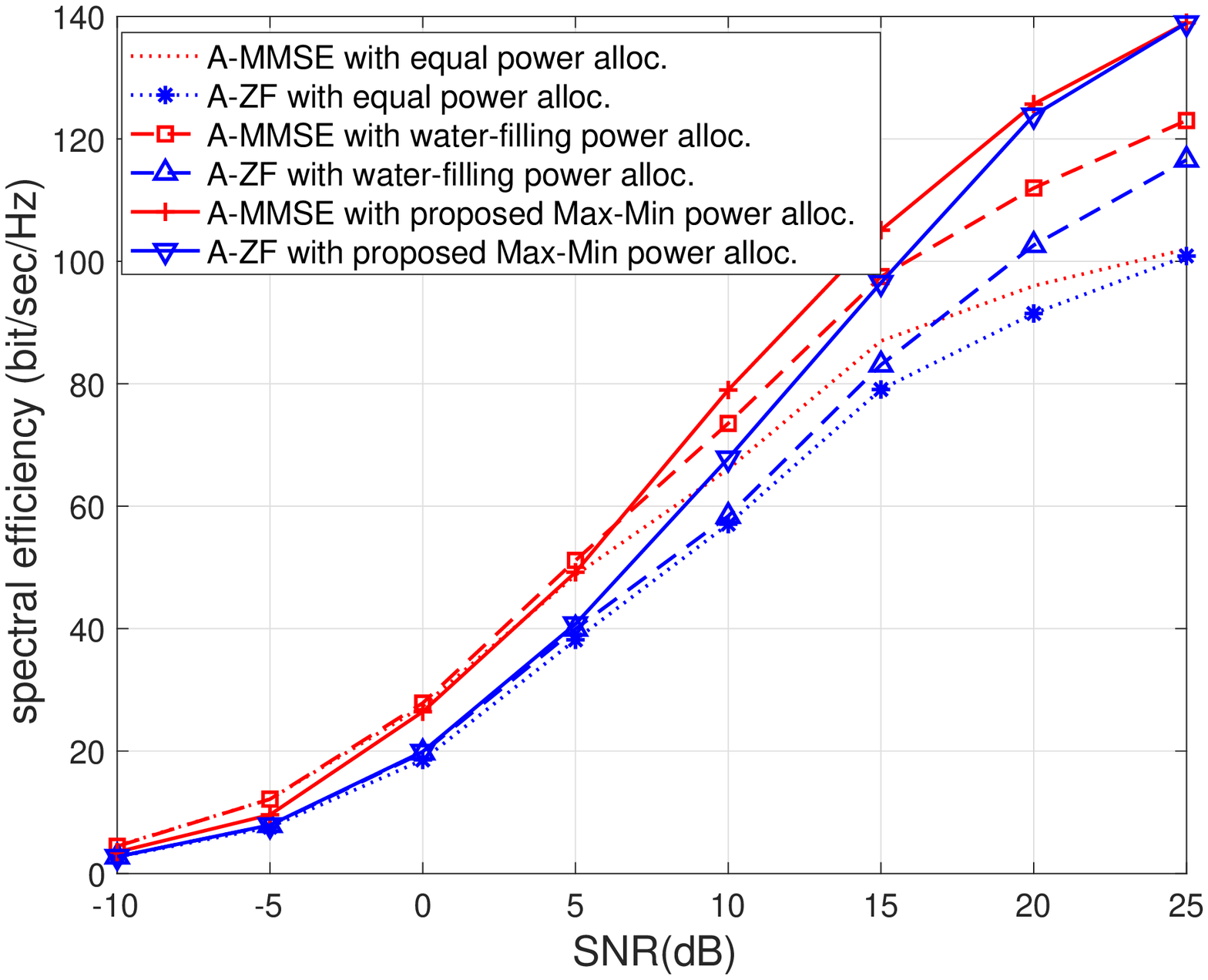}
				\label{fig:sum_Rate_DL_PC}}
			\subfigure[]{			\includegraphics[scale=0.37]{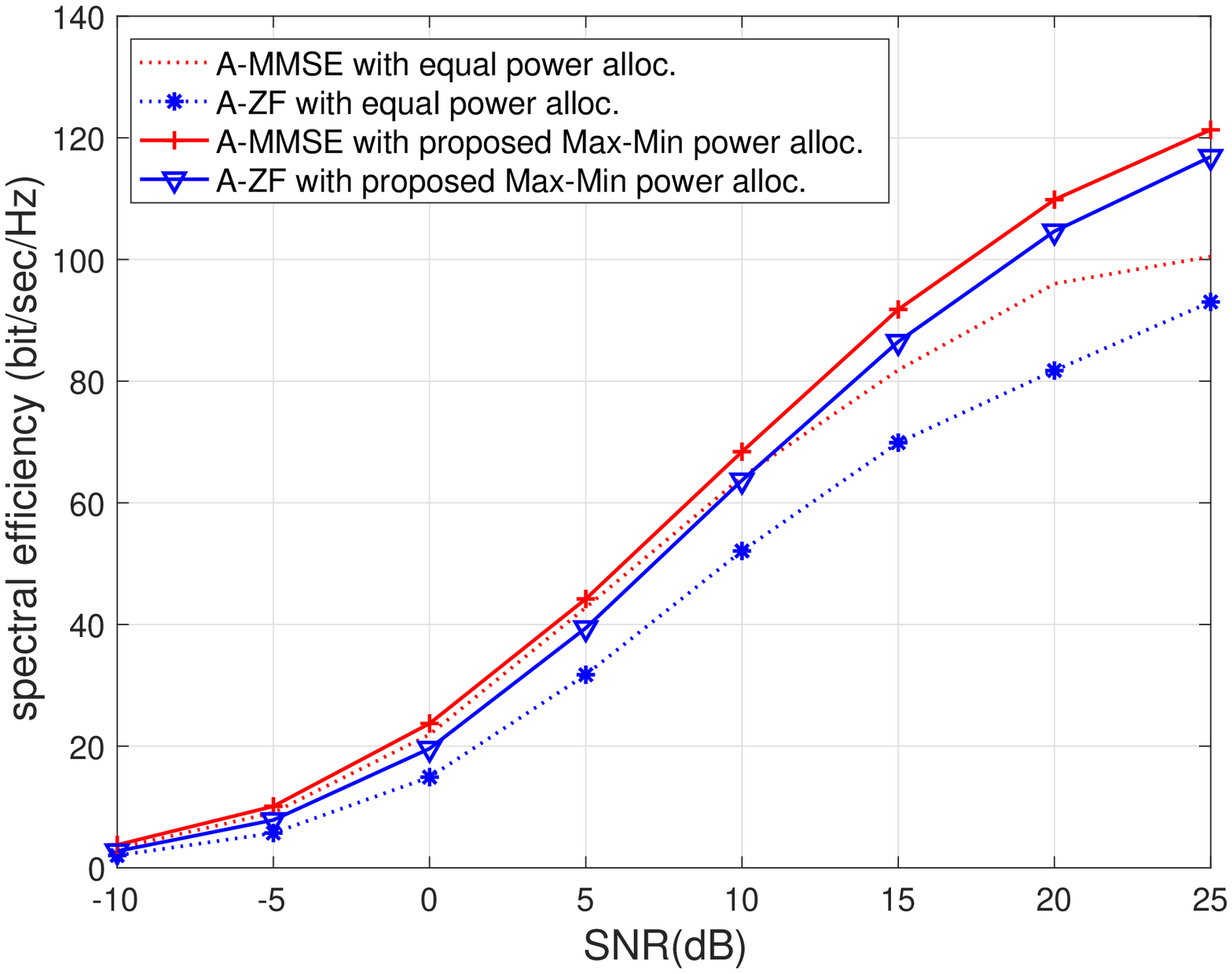}
				\label{fig:sum_Rate_UL_PC}}\vspace{-0.20in}
			\caption{{Spectral efficiency of the proposed combining schemes with equal power control, water-filling power control and the proposed max-min power control versus SNR for $M=10$ APs, and $K=20$ users for the Cell-Free (CF) massive MIMO (AP selection is  not applied): (a) DL and (b) UL.}}
			\label{fig:globfig5}
		\end{figure*}
		\begin{figure*}[t!]
			\centering
			\subfigure[]{
				\includegraphics[scale=0.37]{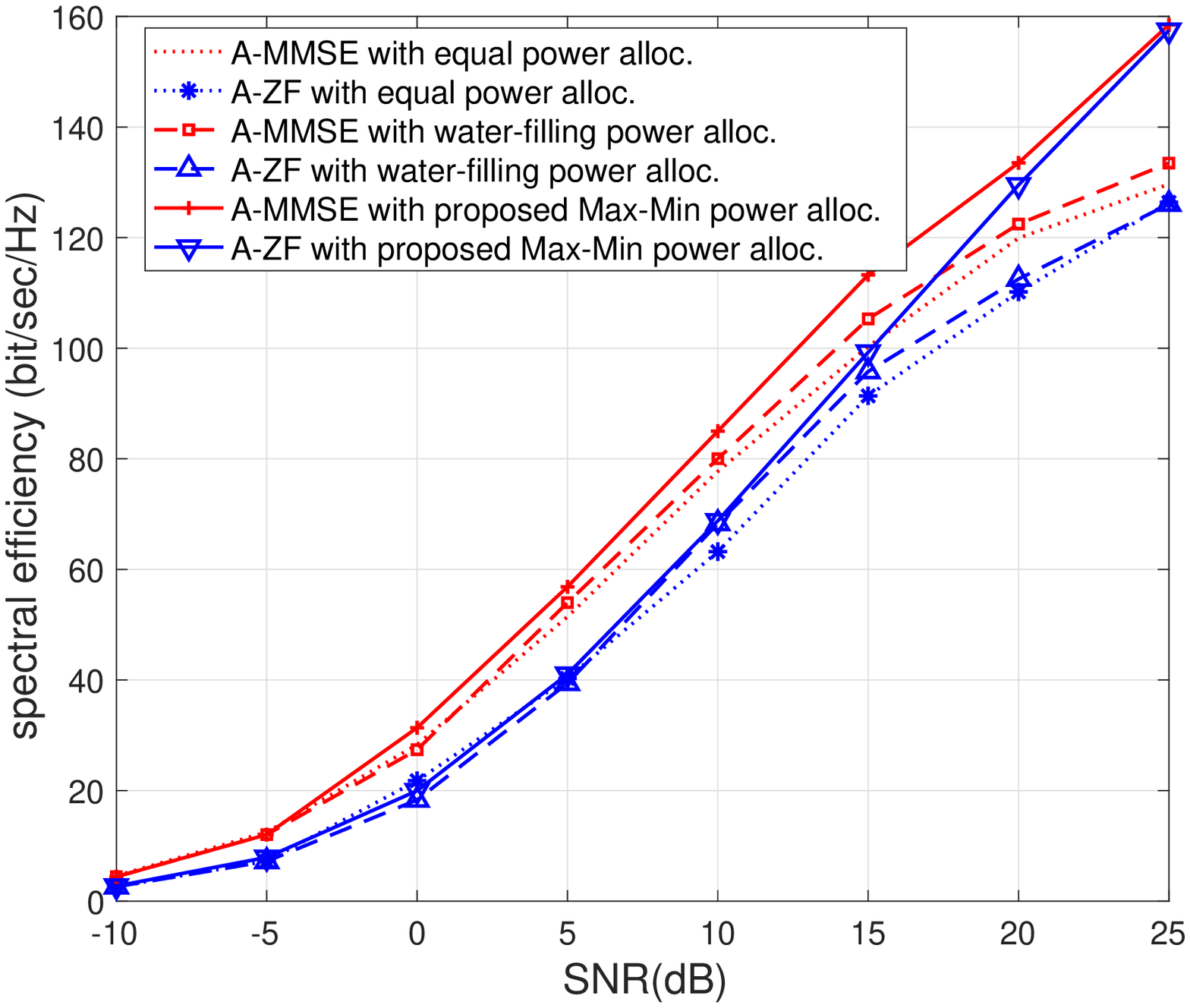}
				\label{fig:sum_Rate_DL_PC_UC}}
			\subfigure[]{
				\includegraphics[scale=0.37]{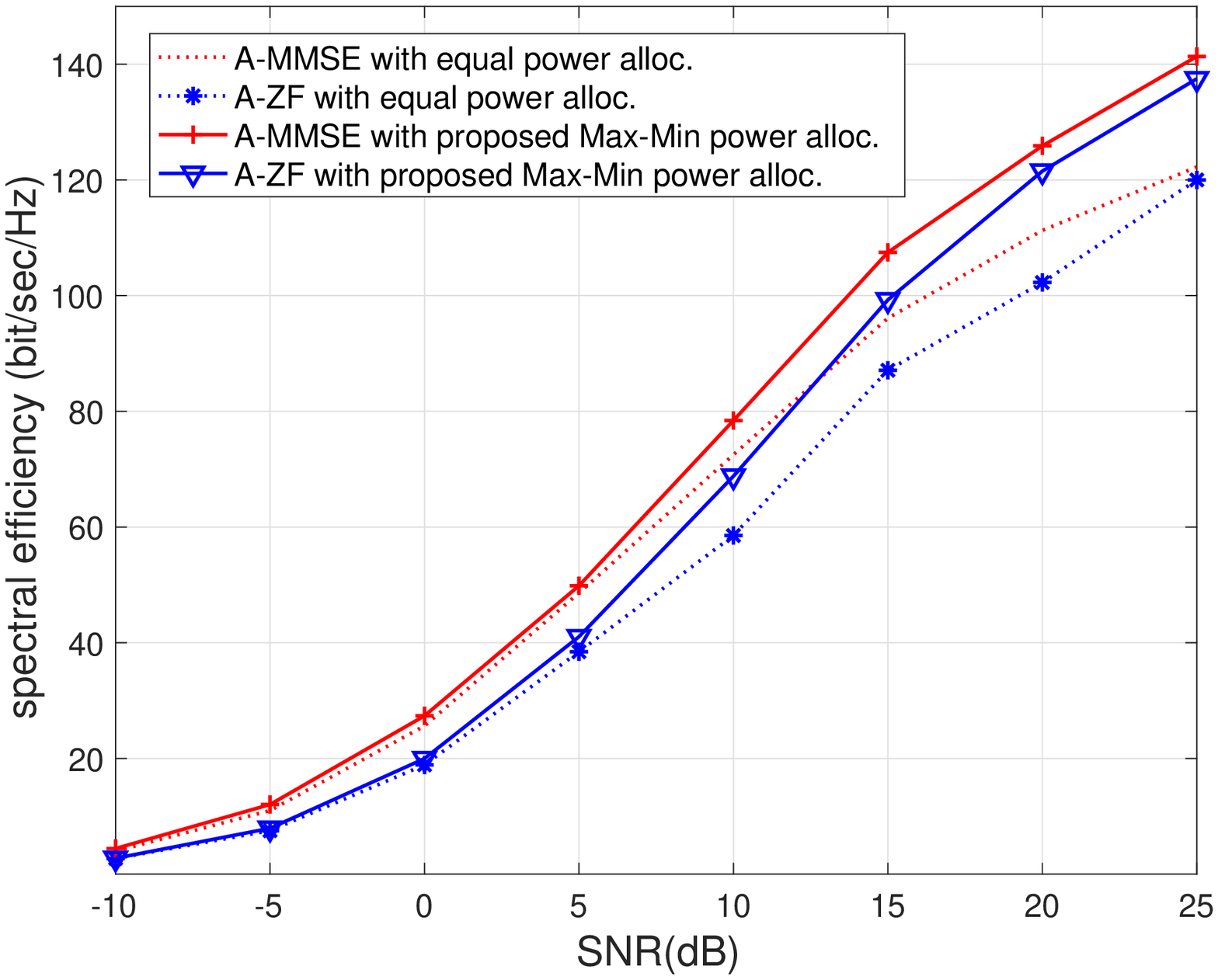}
				\label{fig:sum_Rate_UL_PC_UC}}\vspace{-0.225in}
			\caption{{Same as Fig.~\ref{fig:globfig5} but applying the user centric (UC) AP selection scheme: (a) DL and (b) UL.}}
			\label{fig:globfig6}
		\end{figure*}
		
		\begin{figure*}[t!]
			\centering
			\subfigure[]{
				\includegraphics[scale=0.36]{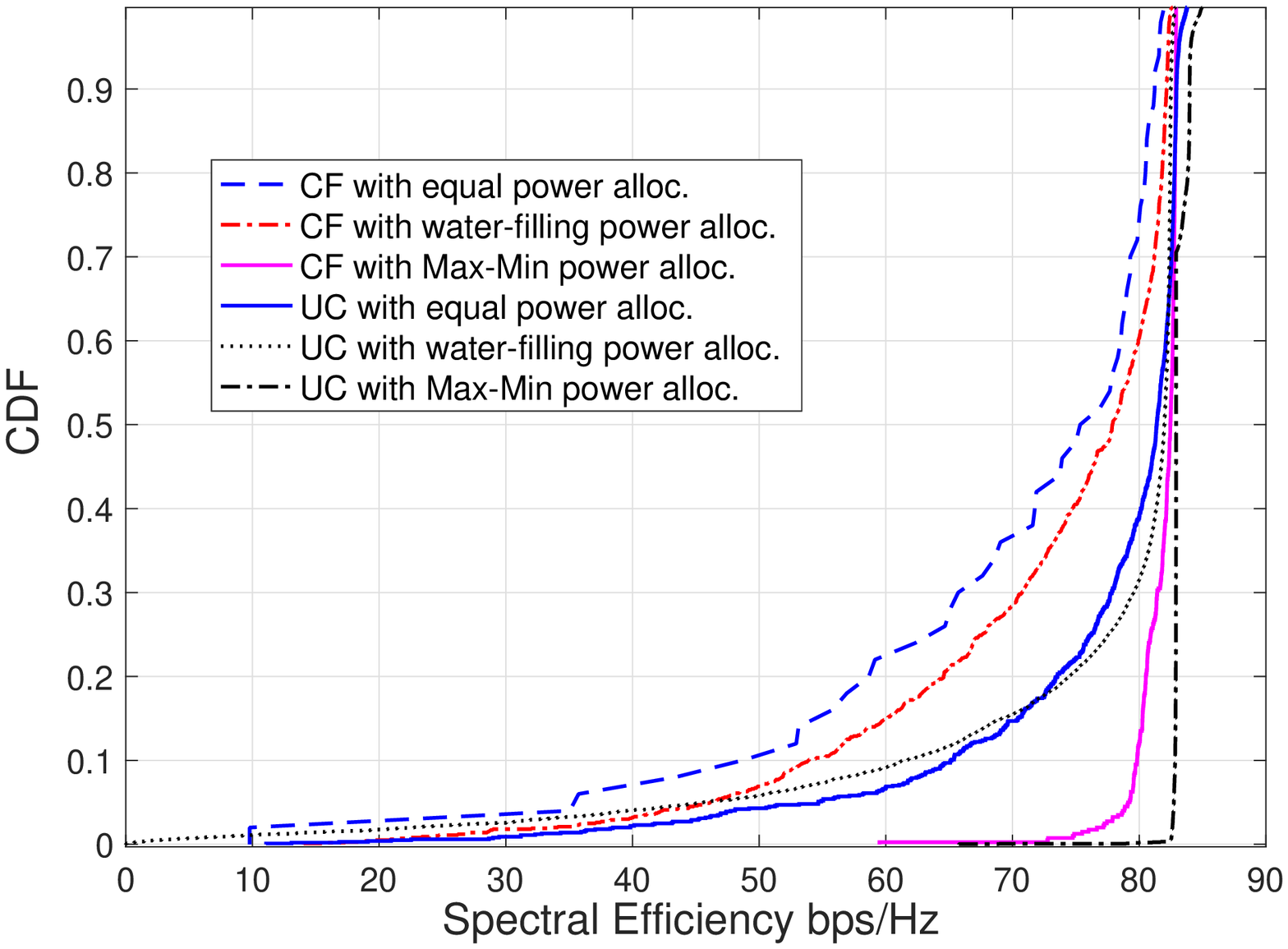}
				\label{fig:CDF_PC}}
			\subfigure[]{
				\includegraphics[scale=0.36]{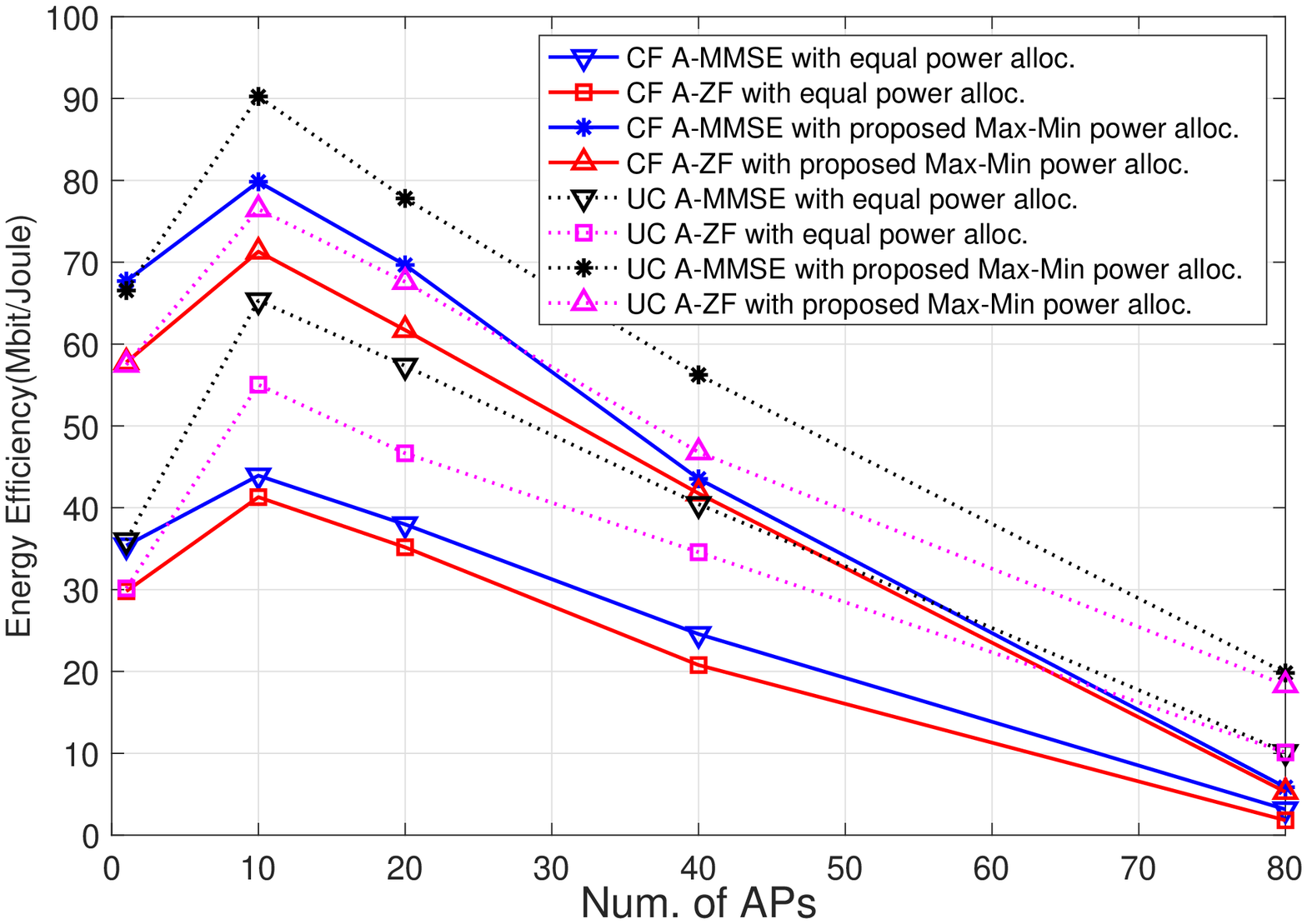}
				\label{fig:EE_DL_PC}}\vspace{-0.15in}
			\caption{{(a) Cumulative distribution of the spectral efficiency for all power control schemes with/without applying the proposed AP selection (CF/UC), and (b) DL energy efficiency of the proposed combining schemes with equal power control and max-min power control versus number of APs. Here, SNR=$\unit[10]{dB}$ for $M=10$, $N=32$, and $K= 20$ users.}}
			\label{fig:globfig7}
			
		\end{figure*}	
		\begin{figure}[t]
			\centering		\includegraphics[scale=0.4]{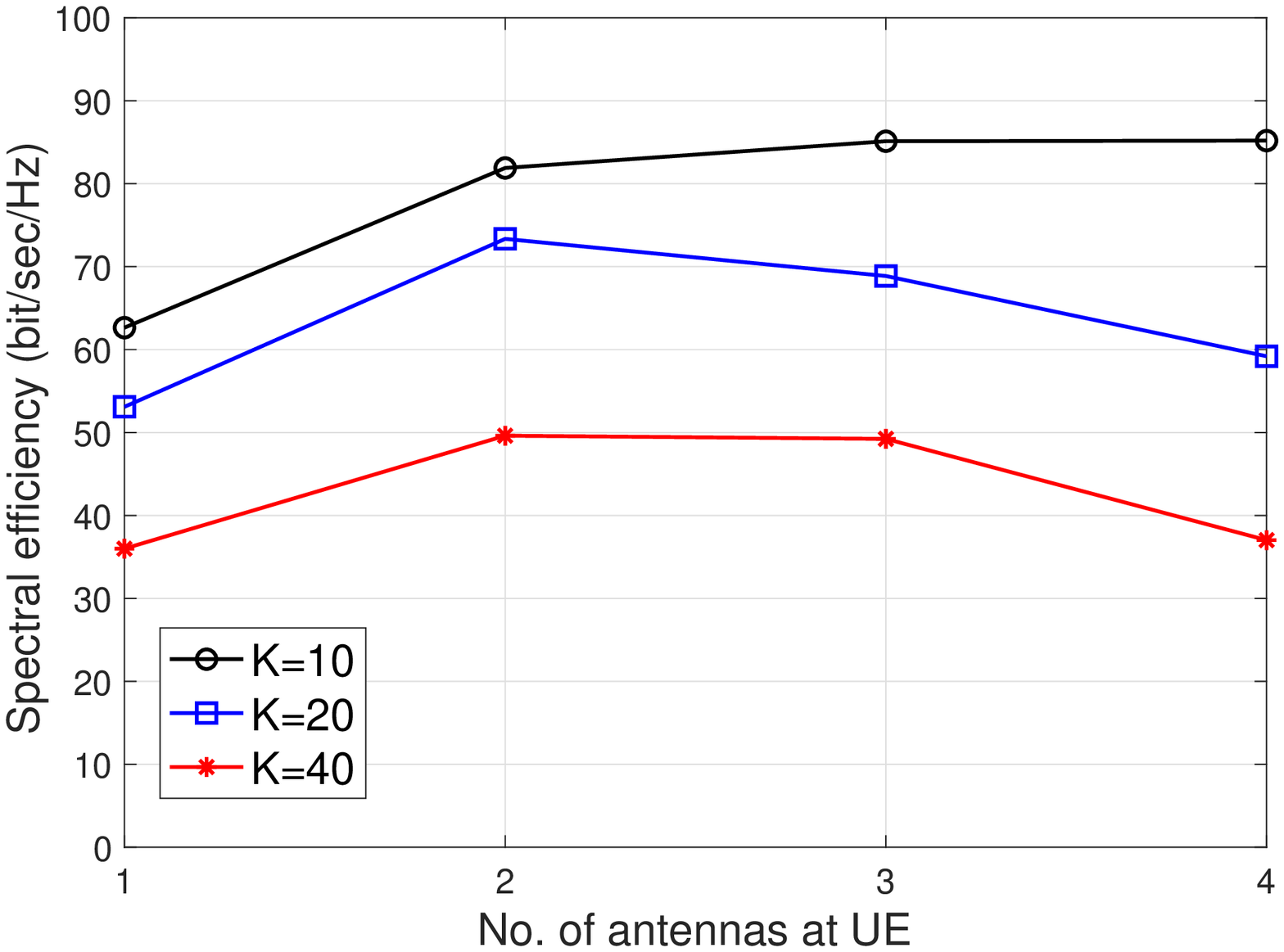}\vspace{-0.15in}
			\caption{{DL spectral efficiency versus multiple antenna configurations at the users for $K=\{10,20,40\}$, $M=10$, and $N=32$.}}
			\label{fig:mult_ant_users}
		\end{figure}
		\subsection{Results and Discussions}
		
		\subsubsection{Performance of Multipath Component Estimation}
		{	In Fig.~\ref{fig:RMSE_theta_beta}, the root mean-square error (RMSE) of the presented multipath component estimation technique is evaluated for $N = 32$ and $T = 16$. We compare the performance of the presented method with that of MUSIC and ESPRIT algorithms, which are subspace-based multipath component estimation techniques that depend on the correlation matrix of the received data \cite{Shmidt1986MUSIC1,Roy1989ESPIRIT} and the gradient-descent-based algorithm \cite{Kim2018CFreeFDD}. The plots demonstrate that the proposed DFT-based technique outperforms the conventional approaches in \cite{Shmidt1986MUSIC1,Roy1989ESPIRIT} and \cite{Kim2018CFreeFDD}. Also, the normalized RMSE performance of the proposed large-scale fading coefficient estimation outperforms that of  conventional subspace-based estimation \cite{Shmidt1986MUSIC1,Roy1989ESPIRIT} and gradient-descent-based estimation \cite{Kim2018CFreeFDD}.  The large scale fading estimation in \cite{Shmidt1986MUSIC1,Roy1989ESPIRIT,Kim2018CFreeFDD} cannot work well when number of samples (snapshots) $T$ is small.}
	
		Moreover, it can be seen that the presented AoA estimation and the large-scale fading estimation method performs slightly worse than that of theoretical bound in ~(\ref{equ:theo_MSE}) since the search grid is large enough ($\mathcal{G}=100$).
		
		\subsubsection{Performance of Spectral Efficiency}
		We compare the performance of the proposed angle-based beamforming and combining schemes (A-MF/A-ZF and A-MMSE) for the FDD-based cell-free massive MIMO with the conventional ideal beamforming and combining schemes (MF/ZF and MMSE) in terms of spectral efficiency for the case of $M=10$ APs with $N=32$ antennas and $K=20$ users. We consider the conventional full-channel-based beamforming and combining schemes (MF/ZF and MMSE) as benchmarks, but they are inapplicable in a realistic FDD cell-free massive MIMO system since complete channel knowledge requires large amount of signaling overhead and feedback.

		For the downlink scenario in Fig. ~\ref{fig:sum_Rate_DLimp}, and for the uplink scenario in Fig.~\ref{fig:sum_Rate_ULimp}, the spectral efficiency of the proposed beamforming/combining schemes with imperfect multipath component estimation is shown. As shown in the figures, the A-MMSE beamforming/combining outperforms A-ZF and A-MF beamforming/combining, due to their ability to suppress interference and noise. In addition, at high SNR (low noise) the A-ZF matches A-MMSE in performance as both of the schemes are able to suppress interference. Moreover at moderate to high SNR values, A-MMSE, A-ZF, and A-MF lead to about $10-40\%$ sum rate loss compared to the conventional ideal beamforming/combining schemes (MF/ZF and MMSE). However, with the proposed angle-based beamforming schemes, the DL CSI signaling overhead is avoided.
		
		Finally, we evaluate the validity of our closed-form expression for the downlink achievable rate for the proposed angle based beamformers given in ~(\ref{equ:Rd1}) with imperfect multipath component estimation. In Fig.~\ref{fig:sum_Rate_DLimp}, we show the accuracy of the proposed closed form of the proposed angle based beamformers ~(\ref{equ:Rd1}) with the simulated form ~(\ref{equ:rate_theo_DL})
		\begin{align}\label{equ:rate_theo_DL}
		\tilde{R}^\mathrm{d}=\sum_{k= 1}^{K}\expec{\log_2 \left( 1+ \frac{\rho^{\mathrm{d}}\sum_{m=1}^{M}{||{\bf h}_{mk}^\mathsf{H}  \hat{\bf w}_{mk}||^{2}}}{\rho^{\mathrm{d}}\sum_{j\neq k}^{K}\sum_{m=1}^{M}{||{\bf h}_{mk}^\mathsf{H}  \hat{\bf w}_{mj}||^{2}}+\sigma_{n}^{2}}\right)}.
		\end{align}
		Moreover, ~(\ref{equ:rate_theo_DL}) represents the achievable rate for genie-aided users that know the instantaneous channel gain \cite{Ngo2017CellFreevsSmallCell}.
		
		In Fig.~\ref{fig:sum_Rate_ULimp}, we also validate the closed-form expression for the uplink achievable rate for the proposed angle based combining given in ~(\ref{equ:Ru1}) for  imperfect multipath component estimation with simulated form~(\ref{equ:rate_theo_UL})
		\begin{align}\label{equ:rate_theo_UL}
		&\tilde{R}^\mathrm{u}=\sum_{k= 1}^{K}\expec{\log_2 \left( 1+ \frac{\rho^{\mathrm{u}}\sum_{m=1}^{M}\!{||{\bf h}_{mk}^\mathsf{H}  \hat{\bf v}_{mk}||^{2}}}{\rho^{\mathrm{u}}\sum_{j\neq k}^{K}\sum_{m=1}^{M}{||{\bf h}_{mk}^\mathsf{H}  \hat{\bf v}_{mj}||^{2}}+\Upsilon_{\sigma}}\right)}.
		\end{align} where $\Upsilon_{\sigma}=\sigma_{n}^{2} \sum_{m=1}^{M}||\hat{\bf v}_{mk}||^{2}$.
		
		One can notice that the closed form achievable rate perfectly matches with Monte Carlo simulated rates. This indicates that our derived expressions ~(\ref{equ:Rd1})  and ~(\ref{equ:Ru1}) are valid performance predictors  of the proposed FDD-based cell-free massive MIMO system.

		\subsubsection{Effect of the Number of APs $M$ for a Fixed Total Number of Service Antennas $(NM)$}
		Furthermore, we examine the performance of the proposed FDD-based cell-free massive MIMO system with different numbers of APs for the downlink case. For fair comparison, the total transmit power in the network is the same, and the number of total service antennas is fixed, i.e. $NM = 320$. {Figure~\ref{fig:sum_Rate_variable_DL_NM} shows the average spectral efficiency ($\kappa \times \sum_k^K R^{\mathrm d}_k$ where $\kappa=1-\frac{\tau}{\tau_c}$, $\tau=K$ corresponds to the length of pilot training sequence in samples,  and $\tau_c$ corresponds to the angle coherence interval in samples) as a function of the number of APs.} We are able to compare the spectral efficiency of cell-free massive MIMO and co-located massive MIMO where the co-located massive MIMO corresponds to the case $M = 1$. It can be seen that the spectral efficiency of the cell-free massive MIMO (for $M=10$ and $N=32$) is better than that of the co-located massive MIMO ($M=1$ and $N=320$) due to spatial diversity gains. However, as the number of APs increases while decreasing the number of antennas per AP, the performance of the cell-free massive MIMO starts to decay. The main reasons for this decay are: 1) for a particular user, there are many APs which are located very far away. These APs will not add significantly to the overall spatial diversity gains which implies that not all APs really participate in serving this user; and 2) angle-based beamforming performs better for higher number of antennas.
		
		\subsubsection{Effect of the Number of Antennas per AP} Finally, to support our findings in Fig.~\ref{fig:sum_Rate_variable_DL_NM}, we study the performance of FDD-based cell-free massive MIMO system with different numbers of antennas per AP for a fixed number of APs ($M=10$) in Fig.~\ref{fig:sum_Rate_variable_DL_M}. As the number of antennas increases, the spectral efficiency increases due to the increased array gain in addition to the applied angle-based beamforming. It can be seen that the spectral efficiency saturates for $N\geq 32$ as no further gains are attained.

		\subsubsection{Performance of the Proposed Power/Weight Control on DL/UL Spectral Efficiency}
		{We compare the DL/UL spectral efficiency performance of the proposed angle-based beamforming and combining schemes (A-ZF and A-MMSE) for the FDD-based cell-free massive MIMO with equal power allocation, water-filling power allocation and the proposed max-min power/weight control for the CF case (AP selection is not applied) and the UC case (AP selection is applied). One can note that the water-filling PC approach is based only on the angle and large-scale fading parameters in which the allocated power is $\rho_{mk}=\max\{\tfrac{1}{\mathcal{K}_m}\left( \rho^{\mathrm{tot}}+\sum_{k\in\mathcal{K}_m} \sigma_n^2(||{\bf A}_{mk}{\bf B}_{mk}||^2)^{-1}\right)-\sigma_n^2(||{\bf A}_{mk}{\bf B}_{mk}||^2)^{-1},0\}$,  where $\rho^{\mathrm{tot}}=\mathcal{K}_m\rho^{\mathrm{d}}$ is the total power, and $\mathcal{K}_m=K$ only if the UC AP selection is not applied. Moreover, the water-filling PC approach is applicable in the DL direction, since only the APs have the knowledge of the angle and large scale fading parameters, whereas for the UL direction the users cannot have this information due to the incurred high signaling overhead.
			
			For the downlink scenario in Figs.~\ref{fig:sum_Rate_DL_PC} and~\ref{fig:sum_Rate_DL_PC_UC}, and for the uplink scenario in Figs.~\ref{fig:sum_Rate_UL_PC} and ~\ref{fig:sum_Rate_UL_PC_UC}, the spectral efficiency using the proposed max-min power/weight control schemes is significantly enhanced compared to the case of equal power control and water-filling power control, especially at high SNR values. In particular, as shown in Fig.~\ref{fig:sum_Rate_DL_PC}, the DL sum-rate of the proposed A-MMSE and A-ZF beamforming using max-min power control is increased by $12\%$-$38\%$ compared to the equal power allocation case. While, in Fig.~\ref{fig:sum_Rate_UL_PC}, the UL sum rate of the proposed A-MMSE and A-ZF combining using max-min weight control is increased by $10\%$-$25\%$ due to the fact that the downlink uses more power (since $\rho^{\mathrm{d}}>\rho^{\mathrm{u}}$) and has more power control coefficients to choose than the uplink does, hence the DL performance is better than the UL performance. Moreover, as shown in  Figs.~\ref{fig:globfig5}, and ~\ref{fig:globfig6}, the UC approach has better performance than that of the CF case since the UEs obtain very noisy signals from the far APs, and not all APs actually participate in serving the users.
			
			In addition,  the cumulative distribution function (CDF) curve for the proposed max-min power control scheme is plotted in Fig.~\ref{fig:CDF_PC}, and compared with the equal PC and the water-filling PC schemes at SNR$=\unit[10]{dB}$. As expected, the max-min PC scheme was able to outperform the rest of the PC schemes and improve the system fairness for both cases CF and UC, respectively. }
		
		\subsubsection{Energy Efficiency versus Number of APs $M$ and a Fixed Total Number of Service Antennas $(NM)$} {Figure~\ref{fig:EE_DL_PC} examines the energy efficiency ~(\ref{EE}) as a function of the number of AP for a fixed total number of service antennas, when the number of AP increases, the number of antennas per AP decreases. As shown, the energy efficiency while applying the proposed max-min power control significantly outperforms that of equal power control by $40\%$-$50\%$, especially when the UC AP selection scheme is applied. Furthermore, we are able to compare the energy efficiency of cell-free massive MIMO and co-located massive MIMO where the co-located massive MIMO corresponds to the case $M = 1$. It can be seen that the energy efficiency of the cell-free massive MIMO (for $M=10$ and $N=32$) is better than that of the co-located massive MIMO ($M=1$ and $N=320$) due to spatial diversity gains, and better spectral efficiency as shown in Fig.~\ref{fig:sum_Rate_variable_DL_NM}. Moreover, the number of APs will affect the level of backhaul power consumption; therefore, as the number of APs increases while decreasing the number of antennas per AP, the performance of the cell-free massive MIMO starts to decay due to the increased backhaul power consumption as shown in ~(\ref{equ:P_backhaul}). }
		
		\subsubsection{Multi-antenna Users extension}
		{	In this subsection, we finally study the effect of having multi-antenna users on the proposed FDD cell-free massive MIMO system where each user is equipped with $N'$ antennas. First, the updated channel model is given by
			\begin{align} \label{mult_ant_channel}
			{\bf H}_{N\times N'}= \sqrt { \frac{1}{ L} }{\bf A}_{}^{\mathrm{AP}} {\bf B}_{} {\bm \Lambda}_{\bm \alpha}({\bf A}^{\mathrm{UE}})^{\mathsf{H}} ,
			\end{align}
			where	${\bf A}_{N\times L}^{\mathrm{AP}}  = [{\bf a}\left (\phi _{1}^{\mathrm{AP}} \right), \dots, {\bf a}\left (\phi _{L}^{\mathrm{AP}} \right)],~  \text{ }{\bf B}_{L\times L} = \mathsf{diag}(\sqrt{\beta_{1}},\dots, \sqrt{\beta_{L}}), $ $({\bm \Lambda}_{\bm \alpha})_{L\times L} = \mathsf{diag}(\alpha_{1},\dots,\alpha_{L}),$ and ${\bf A}_{N'\times L}^{\mathrm{UE}}  = [{\bf a}\left (\phi _{1}^{\mathrm{UE}} \right), \dots, {\bf a}\left (\phi _{L}^{\mathrm{UE}} \right)]$.
			Moreover, the DL spectral efficiency per user is given by
			\begin{align}\label{equ:rate_theo_DL_multi_ant}
			&\kappa \times \!\tilde{R}^\mathrm{d}=\! \left(1-\tfrac{\tau}{\tau_c}\right) \!\times\notag \\&\sum_{k= 1}^{K}\expec{\log_2 \left( \!1+ \!\tfrac{\rho^{\mathrm{d}}\sum_{m=1}^{M}{||\hat{\bf v}_{m^{\star}k}^\mathsf{H} {\bf H}_{mk}^\mathsf{H}  \hat{\bf w}_{mk}||^{2}}}{\rho^{\mathrm{d}}\sum_{j\neq k}^{K}\sum_{m=1}^{M}{||\hat{\bf v}_{m^{\star}k}^\mathsf{H} {\bf H}_{mk}^\mathsf{H}  \hat{\bf w}_{mj}||^{2}}+\sigma_{n}^{2}}\right)},
			\end{align}
			where $\hat{\bf v}_{m^{\star}k}$ corresponds to the combining vector at the multi-antenna $\nth{k}$ user that is based on the estimated AoA of the user from the strongest AP $m^{\star}$. Moreover, the combining vector $\hat{\bf v}_{m^{\star}k}$ follows the same definition as the combining vector defined in Section~ \ref{Sec:Combining} eq. ~(\ref{combining}), but  in this case $\hat{\bf C}_{m^\star}=\hat{\bf A}_{m^{\star}}^{\mathrm{UE}}
			\left((\hat{\bf A}_{m^\star}^{\mathrm{UE}} )^\mathsf{H}\hat{\bf A}_{m^\star}^{\mathrm{UE}} \right)^{-1} $, and the beamforming vector $\hat{\bf w}_{mk}$ follows the same definition as the A-ZF combining vector defined in Section~ \ref{Sec:Beamforming}. The strongest AP $m^{\star}$ is the AP that has the best channel quality with $\nth{k}$ user. One can note that only the $\nth{m^{\star}}$ AP will need to feed back the combining vector $\hat{\bf v}_{m^{\star}k}$ to the $\nth{k}$ user; hence, no extensive signaling overhead is needed from all the APs to feed back the estimated multipath components to the $\nth{k}$ user. Finally, note that $\tau=KN'$ depends on the number of users $K$ and scales linearly with the number of antennas at the users $N'$. Therefore, the factor $(1-\frac{\tau}{\tau_c})$ is an important limiting factor when determining the achievable rates for multi-antenna users.
			
			In Fig.~\ref{fig:mult_ant_users}, the performance of the simulated DL spectral efficiency is studied assuming that $\text{RMSE}_{\hat{\phi^{\mathrm{AP}}}} \!=\!\text{RMSE}_{\hat{\phi^{\mathrm{UE}}}}\!=\!\text{RMSE}_{\hat{\beta}}\!=\!\unit[-18]{dB}$. As shown, the DL spectral efficiency first increases when the number of antennas per user increases. However, this spectral efficiency will reach a peak value and then decrease when the number of antennas per user increases. This is due to the fact that although the spatial diversity per user increases, the multipath channel estimation overhead (the training duration relative to the angle coherence interval) also increases. This channel estimation overhead becomes dominant when $N'$ and $K$ are large.}
		\vspace{-0.1in}
		
		\section{Conclusion}\label{sec:conclusion}
		In this paper, an FDD-based cell-free massive MIMO system that directly acquires multipath components from the uplink pilot signal and processes them for AP cooperation has been considered. It has been shown that an FDD-based cell-free massive MIMO system is a viable alternative compared to a TDD-based system in which angle reciprocity can be exploited to avoid DL CSI feedback and overhead. A low complexity multipath component (AoA and large-scale fading) estimation technique based on DFT operation, along with angle rotation with very small amount of training overhead and feedback cost, has been presented. To evaluate the benefits of the proposed methods, theoretical bounds on the MSE have been derived and validated. In addition, angle-based beamformers and combiners, which incur CSI overhead that scales only with the number of served users rather than the total number of serving antennas, have been proposed. {Finally, a new max-min power/weight control algorithm and associated AP selection scheme that significantly improve the downlink and uplink sum-rate and energy efficiency compared to equal-power allocation and water-filling power control have been proposed.}
		
		{The spectral efficiency of the presented FDD-based cell-free massive MIMO system has been shown to outperform that of cell-based systems for an adequate number of antennas at the APs and a small number of APs.  Furthermore, when the number of active users in the system is small, the spectral efficiency also improves upon equipping the users with an adequate number of antennas.}
		\vspace{-0.1in}

		
		\bibliographystyle{IEEEtran}
		\bibliography{IEEEabrv,bio1}

		\begin{IEEEbiography}[{\includegraphics[width=1in,height=1.25in,clip,keepaspectratio]{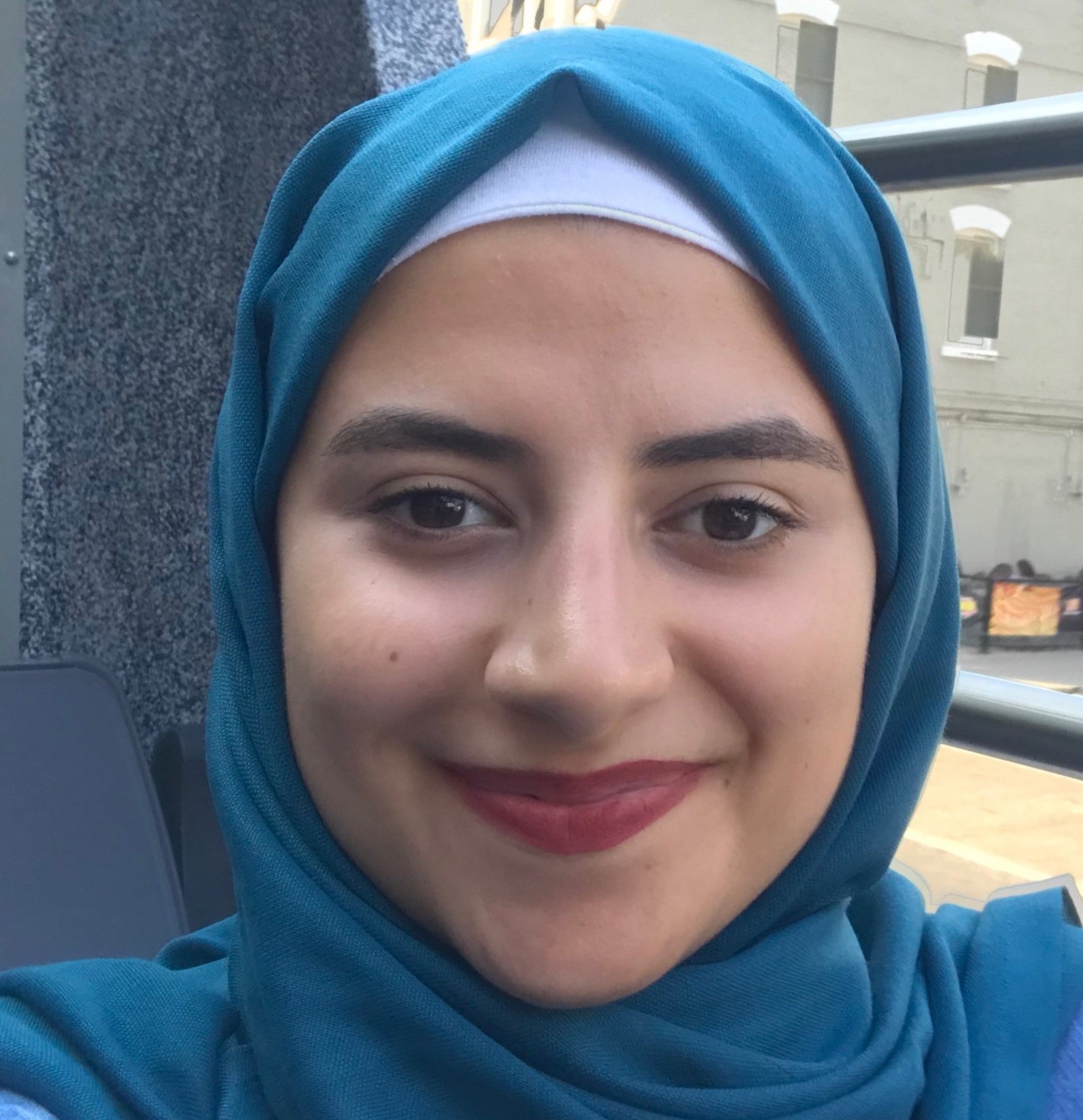}}]{Asmaa Abdallah}  received the B.S. (with High Distinction) and M.S degree in computer and communications engineering from Rafik Hariri University (RHU), Lebanon, in 2013 and 2015, respectively. She is currently pursuing the Ph.D. degree in electrical and computer engineering at the American University of Beirut (AUB), Beirut, Lebanon. 
			She has been a research and teaching assistant at AUB since 2015. She was a research intern at Nokia Bell Labs in France from July 2019 till December 2019, where she worked on new hybrid automatic request (HARQ) mechanisms for long-delay channel in non-terrestrial networks (NTN). She is a current member of the executive committee of IEEE Young Professionals Lebanon’s Section.  Her research interests are in the area of communication theory, stochastic geometry for wireless communications, array signal processing, with emphasis on energy and spectral efficient algorithms for Device-to-Device (D2D) communications, massive multiple-input and multiple-output (MIMO) systems and cell free massive MIMO systems.
			Ms. Abdallah was the recipient of the Academic Excellence Award at RHU in 2013 for ranking first on the graduating class. She also received a scholarship from the Lebanese National Counsel for Scientific Research (CNRS-L/AUB) to support her doctoral studies. 
			
		\end{IEEEbiography}

		\begin{IEEEbiography}[{\includegraphics[width=1in,height=1.25in,clip,keepaspectratio]{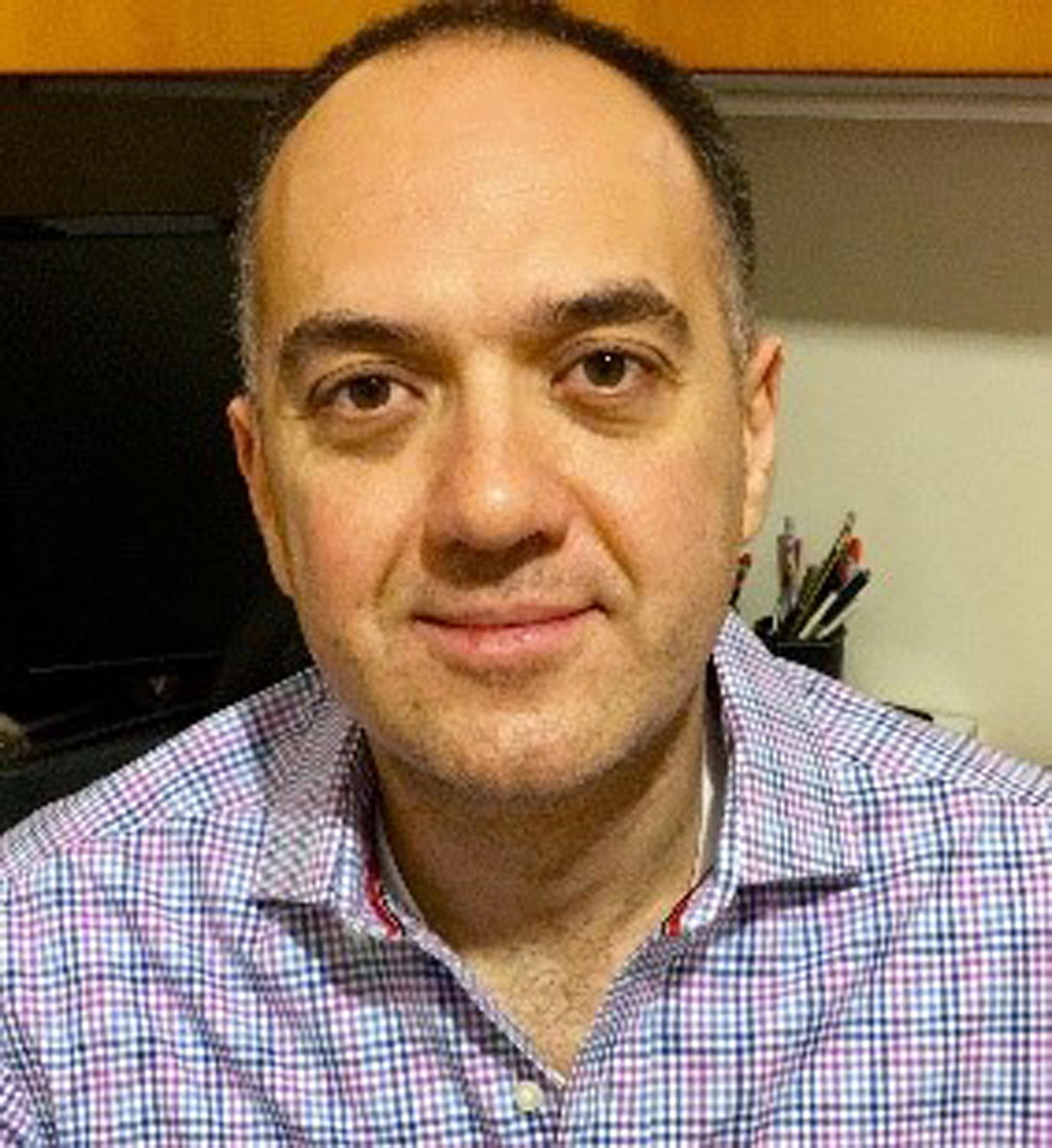}}]{Mohammad M. Mansour}(S'97-M'03-SM'08) received the B.E. (Hons.) and the M.E. degrees in computer and communications engineering from the American University of Beirut (AUB), Beirut, Lebanon, in 1996 and 1998, respectively, and the M.S. degree in mathematics and the Ph.D. degree in electrical engineering from the University of Illinois at Urbana–Champaign (UIUC), Champaign, IL, USA, in 2002 and 2003, respectively.
			
			He was a Visiting Researcher at Qualcomm, San Jose, CA, USA, in summer of 2016, where he worked on baseband receiver architectures for the IEEE 802.11ax standard. He was a Visiting Researcher at Broadcom, Sunnyvale, CA, USA, from 2012 to 2014, where he worked on the physical layer SoC architecture and algorithm development for LTE-Advanced baseband receivers. He was on research leave with Qualcomm Flarion Technologies in Bridgewater, NJ, USA, from 2006 to 2008, where he worked on modem design and implementation for 3GPP-LTE, 3GPP2-UMB, and peer-to-peer wireless networking physical layer SoC architecture and algorithm development. He was a Research Assistant at the Coordinated Science Laboratory (CSL), UIUC, from 1998 to 2003. He worked at National Semiconductor Corporation, San Francisco, CA, with the Wireless Research group in 2000. He was a Research Assistant with the Department of Electrical and Computer Engineering, AUB, in 1997, and a Teaching Assistant in 1996. He joined as a faculty member with the Department of Electrical and Computer Engineering, AUB, in 2003, where he is currently a Professor. His research interests are in the area of energy-efficient and high-performance VLSI circuits, architectures, algorithms, and systems for computing, communications, and signal processing.
			
			Prof. Mansour is a member of the Design and Implementation of Signal Processing Systems (DISPS) Technical Committee Advisory Board of the IEEE Signal Processing Society. He served as a member of the DISPS Technical Committee from 2006 to 2013. He served as an Associate Editor for IEEE TRANSACTIONS ON CIRCUITS AND SYSTEMS II (TCAS-II) from 2008 to 2013, as an Associate Editor for the IEEE SIGNAL PROCESSING LETTERS from 2012 to 2016, and as an Associate Editor of the IEEE TRANSACTIONS ON VLSI SYSTEMS from 2011 to 2016. He served as the Technical Co-Chair of the IEEE Workshop on Signal Processing Systems in 2011, and as a member of the Technical Program Committee of various international conferences and workshops. He was the recipient of the PHI Kappa PHI Honor Society Award twice in 2000 and 2001, and the recipient of the Hewlett Foundation Fellowship Award in 2006. He has seven issued U.S. patents.
		\end{IEEEbiography}
		
	\end{document}